\documentclass[twocolumn,tighten,twocolappendix]{aastex63}

\pdfoutput=1

\usepackage{amsmath, bm, color, colortbl}

\newcommand{\edits}[1]{\textcolor{black}{#1}}
\newcommand{\mdisk}{$M_{\rm d}$}
\DeclareMathAlphabet\mathbfcal{OMS}{cmsy}{b}{n}

\shorttitle{Dynamical Measurements of Disk Masses}
\shortauthors{Andrews et al.}
 
\begin{document}

\title{On Kinematic Measurements of Self-Gravity in Protoplanetary Disks}

\author[0000-0003-2253-2270]{Sean~M.\,Andrews}
\affiliation{Center for Astrophysics \textbar\ Harvard \& Smithsonian, 60 Garden Street, Cambridge, MA 02138, USA}

\author[0000-0002-0786-7307]{Richard~Teague}
\affiliation{Department of Earth, Atmospheric, and Planetary Sciences, Massachusetts Institute of Technology, Cambridge, MA 02139, USA}

\author[0000-0003-1656-011X]{Christopher~P.\,Wirth}
\affiliation{Center for Astrophysics \textbar\ Harvard \& Smithsonian, 60 Garden Street, Cambridge, MA 02138, USA}
\affiliation{Department of Astronomy \& Astrophysics, University of Chicago, 5640 S.~Ellis Avenue, Chicago, IL 60637, USA}

\author[0000-0001-6947-6072]{Jane~Huang}
\affiliation{Department of Astronomy, University of Michigan, 323 West Hall, 1085 S. University Avenue, Ann Arbor, MI 48109, USA}
\affiliation{Department of Astronomy, Columbia University, Mail Code 5246, 538 West 120th Street, New York, NY 10027, USA}

\author[0000-0003-3616-6822]{Zhaohuan~Zhu}
\affiliation{Department of Physics and Astronomy, University of Nevada, Las Vegas, 4505 South Maryland Parkway, Las Vegas, NV 89154, USA}
\affiliation{Nevada Center for Astrophysics, University of Nevada, Las Vegas, 4505 South Maryland Parkway, Las Vegas, NV 89154, USA}

\begin{abstract}
Using controlled injection and recovery experiments, we devised an analysis prescription to assess the quality of dynamical measurements of protoplanetary disk gas masses based on resolved (CO) spectral line data, given observational limitations (resolution, sampling, noise), measurement bias, and ambiguities in the geometry and physical conditions. With sufficient data quality, this approach performed well for massive disks ($M_{\rm d}/M_\ast=0.1$): we inferred \mdisk\ posteriors that recovered the true values with little bias ($\lesssim$\,20\%) and uncertainties within a factor of two (2$\sigma$). The gas surface density profiles for such cases are recovered with remarkable fidelity. Some experimentation indicates that this approach becomes insensitive when $M_{\rm d}/M_\ast\lesssim5$\%, due primarily to degeneracies in the surface density profile parameters. Including multiple lines that probe different vertical layers, along with some improvements in the associated tools, might push that practical boundary down by another factor of $\sim$two in ideal scenarios. We also demonstrated this analysis approach using archival ALMA observations of the MWC 480 disk \citep{maps1}: we measured $M_{\rm d}=0.13^{\: +0.04}_{\: -0.01} \: M_\odot$ (corresponding to $M_{\rm d}/M_\ast=7\pm1$\%) and identified kinematic substructures consistent with surface density gaps around 65 and 135 au. Overall, this (and similar work) suggests that these dynamical measurements offer powerful new constraints with sufficient accuracy and precision to quantify gas masses and surface densities at the high end of the $M_{\rm d}/M_\ast$ distribution, and therefore can serve as key benchmarks for detailed thermo-chemical modeling. We address some prospects for improvements, and discuss various caveats and limitations to guide future work. \\ \\
\end{abstract}

\section{Introduction} \label{sec:intro}

The masses of circumstellar disks are fundamental inputs for all models of planet formation.  The amount of disk material regulates evolution in its structure, which in turn sets key planetary growth efficiencies and drives the early dynamics of planetary systems.  But disk mass (\mdisk) measurements are challenging.  The dominant constituent (cold H$_2$ gas) is inaccessible to observations, and our confidence in the assumed conversions or extrapolations from indirect proxies is limited \citep{bergin18,andrews20,miotello22}.   

The most common \mdisk\ estimates are based on the dust continuum emission at millimeter wavelengths.  If that emission is optically thin, the luminosity is directly proportional to the dust mass \citep{beckwith90}.  The key uncertainties in such conversions are temperatures, opacities, and dust-to-gas ratios: the latter two factors could span large ranges depending on the evolutionary state \citep{birnstiel12,birnstiel16,testi14} and microphysical properties \citep{ossenkopf94,dsharp5} of the dust.  The typical assumptions -- a simple temperature profile and spatially uniform opacity and dust-to-gas ratio, with particle sizes that maximize the opacity -- favor {\it minimal} masses (e.g., \citealt{yliu22}).  In general, an appreciable fraction of the emission is optically thick \citep[e.g.,][]{aw05}, and localized dust over-densities can further bloat the optical depths \citep{ricci12,brogan15,huang18}.  This reinforces that \mdisk\ constraints are lower bounds, since high optical depths hide mass.  Such \mdisk\ underestimates can be severe if the dust albedos are high \citep{zhu19}.      

Although more of an observational challenge, resolved measurements of spectral line intensities offer a more direct probe of the gas reservoir.  The most common diagnostics for estimating \mdisk\ are pure rotational transitions from CO isotopologues \citep{williamsbest14,miotello14,miotello16,zhang17}.  These constraints have analogous systematics to the dust continuum, most significantly due to ambiguities in the molecular abundances relative to H$_2$ \citep{aikawa96,dutrey97,vanzadelhoff01,miotello17,yu17a,schwarz18,krijt20} and high optical depths \citep{booth19}.  The far-infrared HD transitions are appealing alternatives (but currently inaccessible), since the D/H ratio is better understood \citep{bergin13,mcclure16,trapman17}.  But as with CO, \mdisk\ estimates from HD require extrapolations in optical depths, temperatures, vertical structure, and factors that set the local continuum.   

The emerging consensus is that more robust \mdisk\ estimates are accessible by self-consistently modeling a combination of diagnostics, for the dust (multifrequency resolved continuum; e.g., \citealt{carrasco-gonzalez19, macias21, sierra21}) and the gas (resolved groupings of key spectral lines; e.g., \citealt{schwarz16,schwarz21,molyarova17,zhang17,zhang21,anderson19,anderson22,trapman22}).  However, all of these approaches depend on interpreting continuum or line intensities, and are therefore subject to the ambiguities associated with conversion factors from emission to physical conditions and prone to underestimating \mdisk\ (especially for high values) because material is hidden by high optical depths.

A compelling and complementary alternative for gas masses is available in the {\it kinematic} information of spectral line data, through the velocity deviations induced by self-gravity \citep{rosenfeld13a}.  The disk potential increases the gas disk orbital velocities relative to the rate expected from the stellar potential alone, because the total encircled mass is always higher than the stellar mass $M_\ast$.  Though these kinematic signatures are subtle, their interpretation does not rely on line {\it intensities}: systematics associated with conversion factors and optical depths are minimal.  This dynamical method is a new opportunity for accurate \mdisk\ measurements, particularly at high values.  \citet{veronesi21} made the first such measurement, based on Atacama Large Millimeter/submillimeter Array (ALMA) data for the Elias 27 disk.  \edits{\citet{lodato22} and \citet{martire24} then followed up with measurements for the disks in the MAPS project sample \citep{maps1}.}     

These pioneering studies demonstrate the potential of this new, dynamical approach for measuring \mdisk.  But the observational features are subtle, and measurements are fickle: some challenges that complicate the relevant techniques and the interpretation of the results have not yet been fully explored.  This article investigates some of these issues, with a focus on optimizing the technical approach and quantifying the key systematics.  Section \ref{sec:models} designs models of the physical conditions and kinematics expected in disks to serve as `controls' that frame our inquiries.  Section \ref{sec:simobs} outlines the adopted approach for converting those physical models into realistic simulated datasets.  Section \ref{sec:kinematics_concepts} describes how to extract the disk velocity field and infer \mdisk, and Section \ref{sec:results} explores the systematics (accuracy) and precision.  Section \ref{sec:discussion} demonstrates the approach on an archival ALMA dataset and outlines the prospects for deploying what we learned, along with some complications not yet considered in detail.  Section \ref{sec:summary} summarizes the key conclusions.

\section{Disk Models} \label{sec:models}

The primary goal of this article is to explore the precision and systematic uncertainties of dynamical \mdisk\ measurements.  To that end, we designed synthetic datasets where the {\it true} (input) properties of the disks are known, and then investigated how well we could recover those inputs under various conditions.  This section introduces a parametric disk model.  The adopted prescription approximates more sophisticated modeling treatments, but promotes more intuition for factors that complicate dynamical measurements of \mdisk.  The physical conditions, kinematics, and molecular abundances are described in Sections \ref{sec:phys_cond}, \ref{sec:kinematics}, and \ref{sec:abund}, respectively.  A set of fiducial models are assigned in Section \ref{sec:fiducial}, and used to illustrate some key links between structure and dynamics.

\subsection{Physical Conditions} \label{sec:phys_cond}

We defined model structures in cylindrical polar coordinates $(r, \phi, z)$, assuming axisymmetry (no dependence on azimuth $\phi$) and mirror symmetry about the midplane ($z = 0$).  The gas density distribution $\rho$ is defined to be in vertical hydrostatic equilibrium, 
\begin{equation}
    \frac{\partial P}{\partial z} = - \rho \, g_z,
    \label{eq:hse_formal}
\end{equation}
where $P$ is the pressure and $g_z$ is the vertical component of the gravitational acceleration.  Assuming an ideal gas with $P = \rho \, c_s^2$, Eq.~(\ref{eq:hse_formal}) can be written as
\begin{equation}
    -\frac{1}{\rho}\frac{\partial \rho}{\partial z} = \frac{g_z}{c_s^2} + \frac{2}{c_s}\frac{\partial c_s}{\partial z},
    \label{eq:hse}
\end{equation}
where $c_s$ is the sound speed,
\begin{equation}
    c_s = \sqrt{\frac{k_{\textsc{b}} T}{\mu m_{\rm H}}},
    \label{eq:soundspeed}
\end{equation}
with $T$ the temperature, $k_{\textsc{b}}$ the Boltzmann constant, $\mu$ (= 2.37) the mean molecular weight, and $m_{\rm H}$ the mass of a H atom.  The vertical acceleration was decomposed into stellar and disk contributions, $g_z = g_{z, \ast} + g_{z, {\rm d}}$.  The stellar term is the vertical derivative of the potential $\Phi_\ast$,
\begin{equation}
    g_{z, \ast} = \frac{\partial \Phi_\ast}{\partial z} = \frac{G M_\ast z}{(r^2 + z^2)^{3/2}},
    \label{eq:gz_star}
\end{equation}
where $G$ is the gravitational constant.  The disk term is
\begin{equation}
    g_{z, {\rm d}} = 2 \pi G \Sigma,
    \label{eq:gz_disk}
\end{equation}
a simplification \edits{for a mass distribution that is vertically thin and radially infinite \citep{armitage_book}}.  While neither of those assumptions is technically correct, this still serves as a reasonable approximation for our purposes.  Density structures $\rho(r, z)$ were determined by numerical integration of Eq.~(\ref{eq:hse}) for a given $M_\ast$, temperature structure $T(r, z)$, and surface density profile $\Sigma(r)$.

The temperatures are bounded by the midplane ($T_{\rm mid}$) and atmosphere ($T_{\rm atm}$) profiles, using a simplification of the behavior adopted by \citet{dullemond20},
\begin{equation}
    T = T_{\rm mid} + f_T(r, z) \, (T_{\rm atm} - T_{\rm mid}),
    \label{eq:Trz}
\end{equation}
where
\begin{equation}
    f_T(r, z) = \frac{1}{2} \, \tanh{\left[\frac{(z/r) - a_z}{w_z \, a_z}\right]} + \frac{1}{2}.
    \label{eq:Tz}
\end{equation}
The location (midpoint) of the vertical thermal transition is set by $a_z$, and its width is controlled by $w_z$.  The radial temperature profiles
\begin{equation}
    T_{\rm mid} = T_{{\rm m},0} \left(\frac{r}{r_0}\right)^{-q}; \quad 
    T_{\rm atm} = T_{{\rm a},0} \left(\frac{r}{r_0}\right)^{-q}
    \label{eq:Tr}
\end{equation}
define the radial thermal structure.

The surface density profile was defined as an exponentially tapered power-law,
\begin{equation}
    \Sigma = \Sigma_0 \left(\frac{r}{r_0}\right)^{-p} \exp{\left[-\left(\frac{r}{r_t}\right)^{\gamma}\right]}.
    \label{eq:Sigma}
\end{equation}
This prescription is flexible enough to cover the common assumptions in the literature: for example, $\gamma = 2 - p$ reproduces the \citet{lyndenbell74} viscous disk similarity solution, and $\gamma = \infty$ is equivalent to the sharp radial truncation of a power-law at $r_t$.  All parametric normalizations were defined at $r_0 = 10$ au.

\subsection{Kinematics} \label{sec:kinematics}

The bulk motions of the disk models were assumed to be solely in the azimuthal direction, with three contributors to the net velocity distribution, 
\begin{equation}
    v_\phi^2 = \underbrace{r \frac{\partial \Phi_\ast}{\partial r\phantom{\rho}}}_{%
    \textstyle
    \begin{array}{c}
      v_\ast^2
    \end{array}
  } + \underbrace{\frac{r}{\rho}\frac{\partial P}{\partial r}}_{%
    \textstyle
    \begin{array}{c}
      \varepsilon_p
    \end{array}
  } + \underbrace{r \frac{\partial \Phi_{\rm d}}{\partial r \phantom{\rho}}}_{%
    \textstyle
    \begin{array}{c}
      \varepsilon_g
    \end{array}
  }.
    \label{eq:bulk_motion}
\end{equation}
The first term in Eq.~(\ref{eq:bulk_motion}) accounts for the orbital motion in the stellar potential ($\Phi_\ast$), defined as
\begin{equation}
    v_\ast^2 = \frac{G M_\ast r^2}{(r^2 + z^2)^{3/2}} = v_{\textsc{k}}^2 \, \frac{r^3}{(r^2 + z^2)^{3/2}},
    \label{eq:vstar}
\end{equation}
where 
\begin{equation}
    v_{\textsc{k}} = \sqrt{\frac{G M_\ast}{r}} = r \, \Omega_{\textsc{k}}     
\end{equation}
is the Keplerian velocity and $\Omega_{\textsc{k}}$ is the Keplerian angular velocity.  Note that $v_\ast = v_{\textsc{k}}$ at the midplane ($z = 0$).  The second term ($\varepsilon_p$) accounts for radial pressure support, and the third term ($\varepsilon_g$) accounts for the disk self-gravity ($\Phi_{\rm d}$ denotes the disk potential): $\varepsilon_p$ can take positive or negative values, but $\varepsilon_g$ is always positive because the gravitational field ($\partial \Phi_{\rm d}/\partial r$) is an increasing function of $r$.  The stellar potential ($v_\ast$) dominates the velocity field over most of the disk, with the additional two terms present as small deviations.  However, for the diffuse gas reservoir in tapered models (finite $\gamma$) at $r > r_t$, the $\Sigma$ taper implies that $\varepsilon_p$ can control the dynamics.

The deviations $\varepsilon_p$ and $\varepsilon_g$ were calculated numerically, with $\varepsilon_g$ computed following the \citet{bertin99} approximation for the field,
\begin{equation}
    \frac{\partial \edits{\Phi_{\rm d}}}{\partial r} = \frac{G}{r} \int_0^{\infty} \xi(k) \, \sqrt{\frac{r^\prime}{r}} \, k \, \Sigma(r^\prime) \,\, dr^\prime,
    \label{eq:field_eqn}
\end{equation}
with
\begin{equation}
    \xi(k) = \mathcal{K}(k) - \frac{1}{4}\left(\frac{k^2}{1-k^2}\right)\left[\frac{r^\prime}{r}{-}\frac{r}{r^\prime}{+}\frac{z^2}{r r^\prime}\right] \mathcal{E}(k),
    \label{eq:elliptic}
\end{equation}
where $\mathcal{K}$ and $\mathcal{E}$ are complete elliptic integrals of the first and second kind, and $k$ is a composite coordinate 
\begin{equation}
    k = \left[\frac{4 \, r \, r^\prime}{(r + r^\prime)^2 + z^2}\right]^{1/2}.
    \label{eq:k_def}
\end{equation}

For pedagogical purposes, we defined {\it residual} velocity fields that focus on the contributions of pressure support and self-gravity individually and in combination,
\begin{eqnarray}
    \Delta v_p &=& \, \sqrt{v_\ast^2 + \varepsilon_p} - v_\ast \nonumber \\
    \Delta v_g &=& \, \sqrt{v_\ast^2 + \varepsilon_g} - v_\ast  \label{eq:dv} \\ 
    \Delta v_\phi &=& \, v_\phi - v_\ast, \phantom{\sqrt{v_\phi^2}} \nonumber
\end{eqnarray}
that will help serve as visualization aids (e.g., see \citealt{rosenfeld13a}, their Figure 10).

\begin{deluxetable*}{l | c c c c c c c c c c c c c | c c c c c}[t!]
\tabletypesize{\footnotesize}
\tablecaption{Fiducial Model Parameters: fixed $M_{\rm d} = 0.1\,M_\odot$ and $d = 150$ pc \label{table:fid_params}}
\tablehead{
\colhead{} & 
\colhead{$\Sigma_0$} & \colhead{$r_t$} & \colhead{$p$} & \colhead{$\gamma$} & 
\colhead{$T_{\rm m,0}$} & \colhead{$T_{\rm a,0}$} & \colhead{$q$} & \colhead{$a_z$} & \colhead{$w_z$} & \colhead{$M_\ast$} & \colhead{$T_{\rm frz}$} & \colhead{$\sigma_{\rm crit}$} & \colhead{$X_\textsc{co}$} & \colhead{$\Delta x$} & \colhead{$\Delta y$} & \colhead{$i$} & \colhead{$\vartheta$} & \colhead{$v_{\rm sys}$} \\
\colhead{} & \colhead{(g/cm$^{2}$)} & \colhead{(au)} & \colhead{} & \colhead{} & \colhead{(K)} & \colhead{(K)} & \colhead{} & \colhead{} & \colhead{} & \colhead{($M_\odot$)} & \colhead{(K)} & \colhead{(g/cm$^{2}$)} & \colhead{} & \colhead{(\arcsec)} & \colhead{(\arcsec)} & \colhead{(\degr)} & \colhead{(\degr)} & \colhead{(m/s)} 
}
\startdata
\textsf{model A}  & 189 & 75  & 1 & 1 & 40 & 120 & 0.5 & 0.1 & 0.2 & 1.0 & 20 & 0.01 & $10^{-5}$ & 0 & 0 & 30 & 130 & 0 \\
\textsf{model B}  & 100 & 160 & 1 & 2 & 40 & 120 & 0.5 & 0.1 & 0.2 & 1.0 & 20 & 0.01 & $10^{-5}$ & 0 & 0 & 30 & 130 & 0 \\
\textsf{model C}  &  68 & 230 & 1 & 4 & 40 & 120 & 0.5 & 0.1 & 0.2 & 1.0 & 20 & 0.01 & $10^{-5}$ & 0 & 0 & 30 & 130 & 0 \\ 
\enddata
\tablecomments{The reference radius was set to $r_0 = 10$ au.}
\end{deluxetable*}

\subsection{Molecular Abundances} \label{sec:abund}

As noted in Section \ref{sec:intro}, the abundance distribution for the molecule of interest is a crucial factor in converting physical conditions into line emission.  The focus here is CO.  Motivated by astrochemical models \citep{aikawa96,aikawa02}, we set a constant abundance $X_\textsc{co}$ in a vertical layer (cf., \citealt{qi08}).  The lower boundary of that layer is defined as the isotherm where CO adsorbs onto dust grains: $T = T_{\rm frz}$ \citep{vanzadelhoff01}.  The upper boundary is the surface $z_{\rm crit}(r)$ associated with a critical column density $\sigma_{\rm crit}$,
\begin{equation}
    \sigma_{\rm crit} = \int_\infty^{z_{\rm crit}} \rho \, dz, 
    \label{eq:scrit}
\end{equation}
meant to emulate the photodissociation threshold  \citep{vanzadelhoff03,aikawa06}.  

We defined the abundance structure as
\begin{equation}
    \rho_\textsc{co} = D_\textsc{co} \, X_\textsc{co} \, \rho,
\end{equation}
where the depletion factor for regions outside the abundant layer is a step-function 
\begin{equation}
    D_\textsc{co} = 
\begin{cases}
    1 & \text{if } T \geq T_{\rm frz} \text{ and } z \leq z_{\rm crit}, \\
    0.001              & \text{otherwise}.
\end{cases}
\label{eq:abund}
\end{equation}
We adopted $X_\textsc{co} = 10^{-5}$ in these models, roughly the mean value measured by \citet{zhang21} and predicted in detailed evolution models by \citet{krijt20}.

\subsection{Fiducial Models} \label{sec:fiducial}

We designed three fiducial disk models to guide an exploration of the quality of \mdisk\ inferences, with the parameters compiled in Table \ref{table:fid_params}.  These models all consider a $M_\ast = 1\,M_\odot$ star hosting a $M_{\rm d} = 0.1\,M_\odot$ disk, but with different $\Sigma(r)$ behaviors at large $r$ that permit us to focus on variations associated with regions where the dynamical signal of self-gravity is strongest.

\begin{figure}
    \centering
    \includegraphics[width=\linewidth]{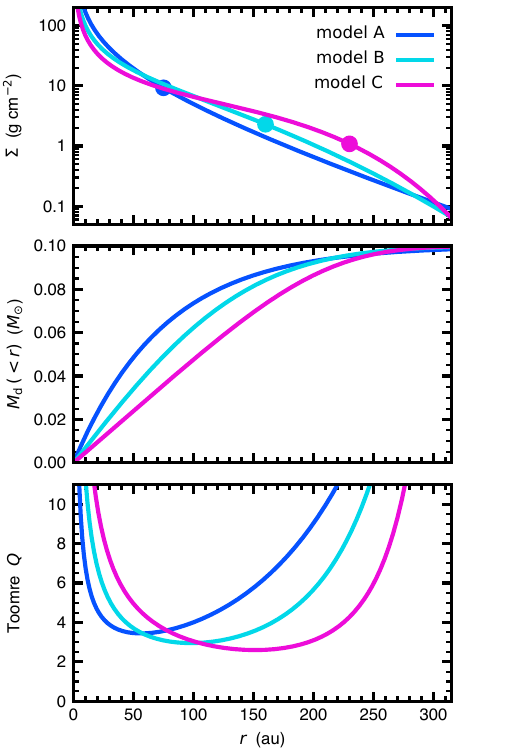}
    \caption{The surface density \edits{(circles mark $r_t$)}, encircled mass, and Toomre stability parameter radial profiles for the fiducial models.  Note that these models are formally stable, even though the canonical mass ratio ($M_{\rm d} / M_\ast = 0.1$) is high.}
    \label{fig:sigma_profiles}
\end{figure}

Figure \ref{fig:sigma_profiles} shows radial profiles of $\Sigma$, encircled mass 
\begin{equation}
    M_{\rm d}(<r) = \int_0^r \Sigma \, 2 \pi r^\prime dr^\prime,
    \label{eq:Mencircled}
\end{equation}
and the \citet{toomre64} instability parameter
\begin{equation}
    Q = \frac{c_s \Omega_{\textsc{k}}}{\pi G \Sigma}
    \label{eq:toomre}
\end{equation}
for each model.  These models are formally gravitationally stable ($Q > 1$), with minimum $Q$ values $\approx$\,3.  The key variation between models is in the $\Sigma$ taper ($\gamma$): $p$ was fixed, $r_t$ was adjusted to force the models to have the same effective emission size (Section \ref{sec:simobs}).  

\begin{figure}[t!]
    \centering
    \includegraphics[width=\linewidth]{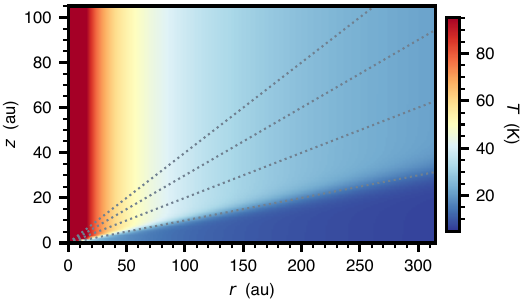}
    \caption{The temperature structure for the fiducial models.  Dotted curves mark $z/r = 0.1$, 0.2, 0.3, and 0.4.}
    \label{fig:Trz}
\end{figure}

\begin{figure}[t!]
    \centering
    \includegraphics[width=\linewidth]{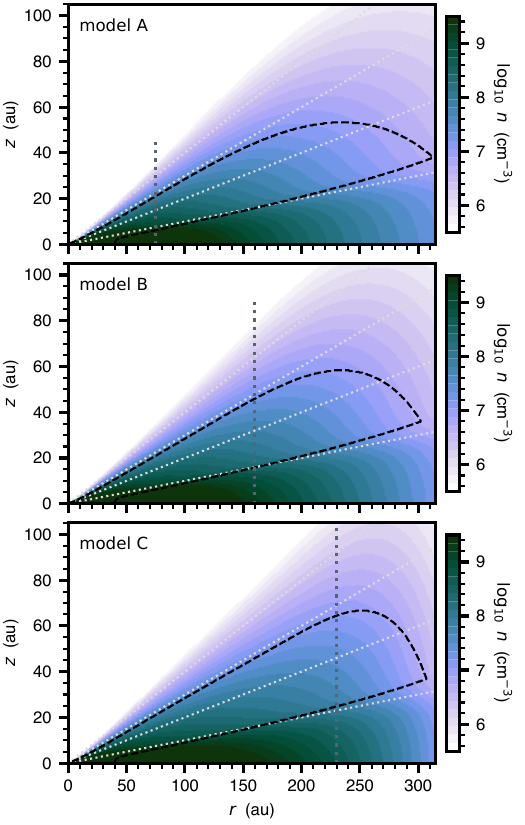}
    \caption{The (volume) density structures (on a $\log_{10}$ scale) for the fiducial models.  The dashed contour regions \edits{enclose the CO-rich layer, and the dotted vertical lines mark $r_t$}.  Other annotations are as in Figure \ref{fig:Trz}.}
    \label{fig:nrz}
\end{figure}

The thermal structures for the fiducial models are identical: Figure \ref{fig:Trz} shows their $T(r, z)$ distribution.  The associated parameters were chosen to emulate \edits{the predictions from} more sophisticated radiative transfer models\edits{, particularly for scenarios where modest dust settling (less turbulent stirring) introduces a steeper vertical temperature gradient at lower altitudes} \citep[e.g.,][]{dalessio06}.  Figure \ref{fig:nrz} shows the volume density distributions for each model (where $n = \rho / \mu m_{\rm H}$) and marks the layer where CO is abundant in the gas-phase, bounded between the $T_{\rm frz}$ isotherm and $z_{\rm crit}$ surface.  These layers span $z/r \approx 0.1$--0.3 outside the midplane CO snowline ($r = 40$ au), consistent with ALMA measurements \citep{law21,law22,rich21,izquierdo21,paneque-carreno23}.  \edits{Note that the ``pinch" in the $n(r, z)$ distribution near $z/r = 0.1$ is a consequence of the steep transition in the adopted $T(z)$ prescription.}

\begin{figure*}[t!]
    \centering
    \includegraphics[width=\linewidth]{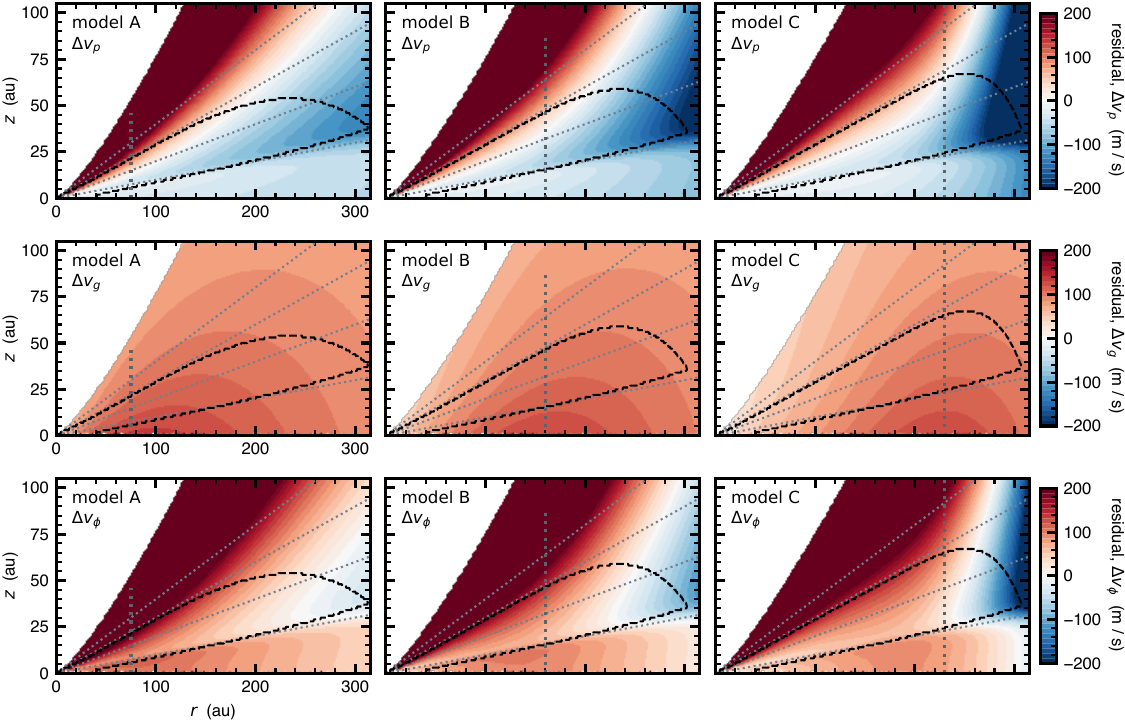}
    \caption{The residual velocity structures (Eq.~\ref{eq:dv}) for the fiducial models (linear scale).  The annotations are as in Figure \ref{fig:nrz}.}
    \label{fig:dv}
\end{figure*}

Figure \ref{fig:dv} highlights the residual velocity fields (Eq.~\ref{eq:dv}).  The $\Delta v_p$ contributions exhibit a vertically layered structure, where deviations near the base of the CO-rich layer are slightly ($|\Delta v_p|/v_\ast \sim$ 1\%, or $|\Delta v_p| \approx 10$--30 m s$^{-1}$) negative before diverging to much stronger negative deviations for $r \gg r_t$ (in the $\Sigma$ taper, where $\varepsilon_p$ dominates).  The flared density distribution reverses the sign of the radial pressure gradient at higher altitudes, transitioning to positive $\Delta v_p$ residuals for $z / r \gtrsim 0.25$ (or higher in the $\Sigma$ taper).  In contrast, $\Delta v_g$ increases with $r$ until $\Sigma$ tapers (where the disk potential gradient diminishes) and is always positive.  The largest $\Delta v_g$ are at the midplane, but the vertical variation is modest.

\begin{figure*}
    \centering
    \includegraphics[width=\linewidth]{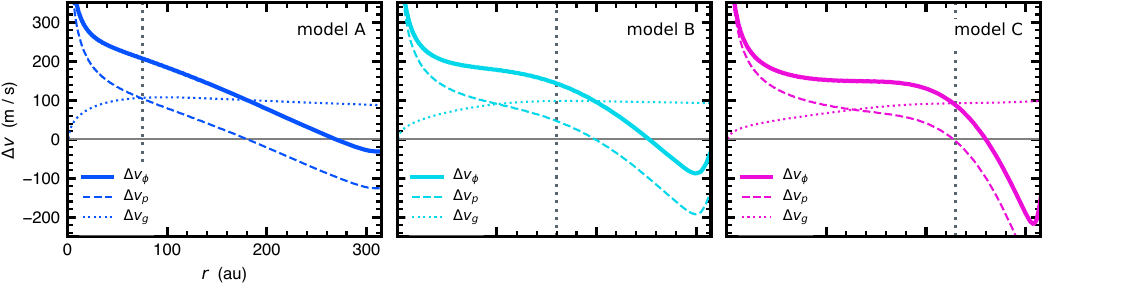}
    \caption{Decompositions of the velocity residuals (Eq.~\ref{eq:dv}) for each model along the top boundary of the CO-rich layer (see Fig.~\ref{fig:dv}), which coincides with the CO emission surface (see Sect.~\ref{sec:res_bias}).  \edits{The vertical lines mark $r_t$ as a point of reference.} \\ \\}
    \label{fig:dv_decompositions}
\end{figure*}

Assuming that the top of the CO-rich layer coincides with the spectral line ``photosphere" (see Sect.~\ref{sec:res_bias}), Figure \ref{fig:dv_decompositions} summarizes the key behaviors of the observable residual velocities.  We see the kinematic deviations from pressure support dominate at both small and large $r$: self-gravity is a modest contributor when a small fraction of the disk mass is encircled, and the pressure gradient becomes very strong well out in the $\Sigma$ taper.  The contributions of pressure support and self-gravity are comparable ($\Delta v_p{\sim}\Delta v_g$) at radii within a factor of $\sim$two around $r_t$.  The transition between these regimes is determined by the $\Sigma$ gradient ($\gamma$) in the taper (which motivated the fiducial parameter choices).

\section{Simulated Data} \label{sec:simobs}

ALMA observations of molecular emission lines can be powerful probes of disk dynamics \citep{pinte22}.  The motions of interest here are small perturbations on the ``background" field, with $|\Delta v_\phi| / v_\ast \lesssim 5$\%.  The precision at which such subtle features are measured depends critically on the sensitivity in the selected tracer \citep[e.g.,][]{teague22}, so observers have focused on bright, optically thick line emission from abundant species.  The discussion here is focused on the $J$=2$-$1 pure rotational transition of CO, though we also address some benefits of extending this analysis to other transitions, isotopologues, and species.  This section describes how we translated the physical and dynamical conditions of the fiducial models into (synthetic) observations (Sect.~\ref{sec:RT}), and  explains how the kinematic perturbations of interest are manifested in such datasets (Sect.~\ref{sec:cubes}).

\subsection{Model Processing to Simulated Datasets} \label{sec:RT}

Using the prescription we described above and the parameters in Table \ref{table:fid_params}, we relied on the (local thermodynamic equilibrium) excitation and raytracing capabilities of {\tt RADMC-3D} \citep{dullemond12} to compute synthetic spectral line cubes.  The model properties were assigned on a fixed grid in spherical coordinates (some nuances of that procedure are detailed in Appendix \ref{app:grid}).  The CO quantum properties were compiled from the {\sc LAMDA} database \citep{schoier05}.  Line profiles were assumed to be Gaussian with widths set solely by thermal broadening (see \citealt{flaherty18}).    

\begin{figure*}[ht!]
    \centering
    \includegraphics[width=\linewidth]{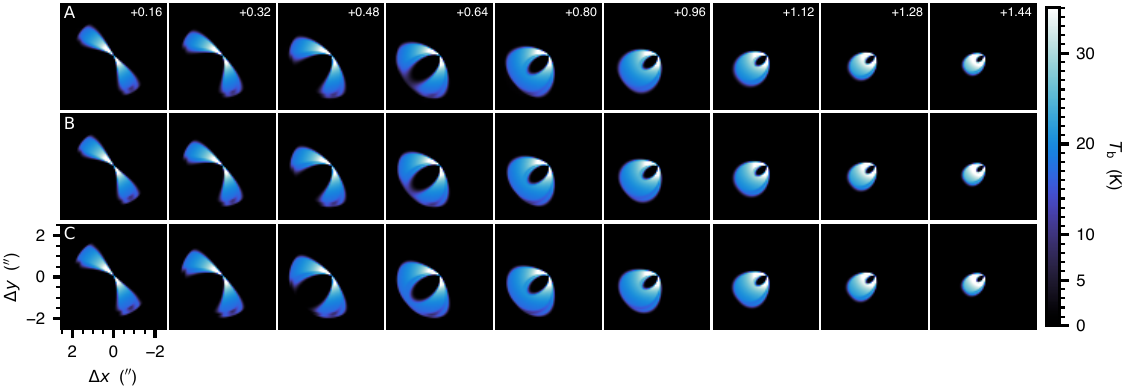}
    \caption{A representative subset of the \textit{\textbf{raw}} channel maps for the fiducial models (see Table \ref{table:fid_params}), on the same (Rayleigh-Jeans) brightness temperature scale.  This shows a portion of the redshifted side of the line, with the LSRK velocities (in km s$^{-1}$) noted in the top right corners of the panels in the top row.  \edits{These maps are showing only every other 80 m s$^{-1}$ channel, to illustrate the behavior over a broader projected velocity range.}  Each panel corresponds to a 750 au field of view. \\}
    \label{fig:chanmap_models}
\end{figure*}

Some additional extrinsic quantities are used in the raytracing (see Table \ref{table:fid_params}).  The observing geometry is determined by the distance $d$ and the sky-projected inclination $i$ and position angle $\vartheta$.  We assigned $d = 150$ pc, $i = 30\degr$, and $\vartheta = 130\degr$, using the convention where $\vartheta$ is the angle E of N to the redshifted side of the major axis, and set the systemic velocity $v_{\rm sys}$ and disk center ($\Delta x$, $\Delta y$) to the origins of their respective reference frames (formally, $\Delta x = \Delta \alpha \cos{\delta}$ and $\Delta y = \Delta \delta$, where \{$\Delta \alpha$, $\Delta \delta$\} are the R.A. and Declination offsets from the observed pointing).  The {\tt RADMC-3D} raytracing was configured to produce a model cube on specified spatial and spectral grids.  We adopted 10 milliarcsecond pixels in a 1024$\times$1024 ($\sim$10\arcsec) field of view and channels with a native spacing of 30.5 kHz ($\sim$40 m s$^{-1}$ at the $J$=2$-$1 rest frequency) that span $\pm$6.5 km s$^{-1}$ around $v_{\rm sys}$.  

We designed a template for ALMA observations of the CO $J$=2$-$1 line (rest frequency 230.538 GHz) that emulates the kind of program we expect to be most fruitful for dynamical \mdisk\ measurements.  It comprises a group of five 1 hour execution blocks (EBs): one with the C--4 configuration (15--780 m baselines), and four with the C--7 configuration (64--3600 m baselines).  We assumed these occur during Cycle 11, with observations in C--4 on 2025 March 1 (HA = $\pm$0.5$^{\rm h}$), and C--7 on 2025 May 20 and 25 (both with consecutive EBs at HA ranges from $-1$ to 0$^{\rm h}$ and 0 to 1$^{\rm h}$).  For each EB, we generated a set of Fourier ($u$, $v$) coordinates for each (pre-averaged) 18 s timestamp, with an associated grid of native channels in the kinematic local standard of rest (LSRK) frame appropriate for the time, date, and source location (R.A.=16$^{\rm h}$, Decl.=$-$30\degr) for a topocentric grid that was fixed at the initial timestamp.

\begin{figure*}[ht!]
    \centering
    \includegraphics[width=\linewidth]{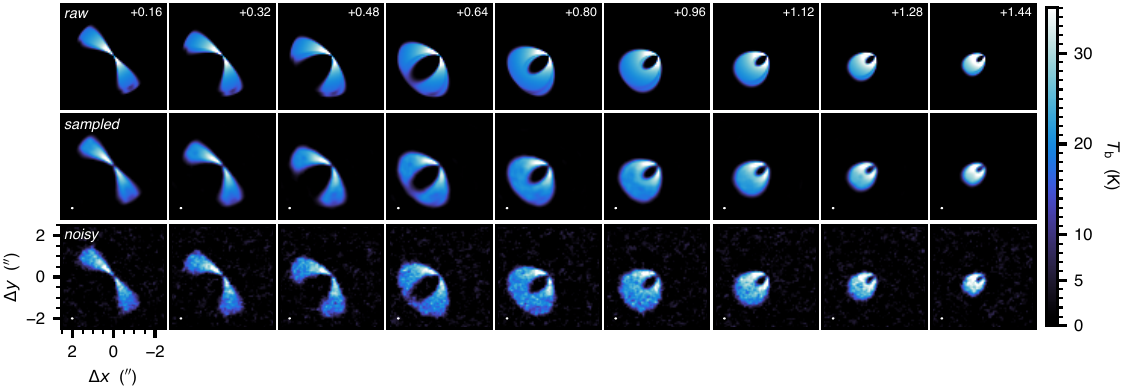}
    \caption{The same subset of channel maps (and annotations, colorscale) as in Figure \ref{fig:chanmap_models} for {\sf model B}, now showing (from top to bottom rows) the \textit{\textbf{raw}}, \textit{\textbf{sampled}}, and \textit{\textbf{noisy}} versions of the cube (note the top row here is identical to the second row in Figure \ref{fig:chanmap_models}).  The FWHM synthesized beam dimensions are shown as (small) white ellipses in the bottom left corners of the sampled and noisy channel maps (the effective pixel resolution in the raw channel maps is $\sim$10$\times$ smaller). \\}
    \label{fig:chanmap_cubetypes}
\end{figure*}

A model cube was created at the LSRK frequencies corresponding to the topocentric channels at the midpoint of each EB.  Their Fourier transforms were sampled onto the template ($u$, $v$) spatial frequencies using the {\tt vis\char`_sample} package \citep{loomis18}.  The visibility spectra were interpolated to the appropriate LSRK frequencies at each timestamp.  These {\it pure} visibility spectra were converted into two datasets.  For one, we convolved the pure spectra with the \edits{default} ALMA spectral response function \edits{(the Fourier transform of the Hann window function)}.  For the other, we injected noise into the pure spectra \edits{using random draws from a Gaussian distribution with mean 0 and standard deviation scaled to produce the RMS expected (2.3 mJy beam$^{-1}$ per 80 m s$^{-1}$ channel) for the integration time (based on the online ALMA exposure time calculator\footnote{\url{https://almascience.nrao.edu/proposing/sensitivity-calculator}})}, and then convolved with the spectral response function.  Finally, the datasets for all EBs were concatenated.  The simulations and post-processing steps were computed with the {\tt csalt} package (Andrews et al., {\it in preparation}).       

These two datasets were then converted into their respective `observed' cubes with the standard imaging process using the {\tt CASA/tclean} task \citep{CASA22,cornwell08}.  We set 0\farcs02 pixels and a channel spacing of 80 m s$^{-1}$ (matching the spectral resolution) spanning $v_{\rm sys} \pm 5.6$ km s$^{-1}$, and used Briggs weighting with {\tt robust} = +1 and {\tt multiscale} parameters that included point source and Gaussian scales at 10, 30, and 50 pixel FWHMs.  These cubes were {\tt clean}ed to a threshold of \edits{4.6 mJy beam$^{-1}$ (or 7.4 K, twice the noise level)} using a Keplerian mask \citep{teague20_kepmask} that was convolved with a FWHM = 0\farcs15 Gaussian.  They have synthesized beams with FWHM = $0\farcs13\times0\farcs11$ (at P.A. = 95\degr).  

Three versions of a cube are produced for each model.  The \textit{\textbf{sampled}} cube treats the spatial and spectral sampling of ALMA observations, in the absence of noise.  The \textit{\textbf{noisy}} cube is the version of the sampled cube with realistic noise.  And the \textit{\textbf{raw}} cube was computed as the direct output of {\tt RADMC-3D} raytracing on these {\it imaged} velocity channels at the full (0\farcs01) resolution.  The raw cube is intended as a reference for probing the effects of line radiative transfer on inferred disk properties without observational sampling issues or noise.  

Figure \ref{fig:chanmap_models} highlights a representative subset of channel maps from the raw cubes for the fiducial models, to illustrate their general behavior and subtle differences.  Figure \ref{fig:chanmap_cubetypes} compares the three different types of cubes -- raw, sampled, and noisy -- using {\sf model B} as an example.

\subsection{Dynamical Effects on the Cubes} \label{sec:cubes}

While the fiducial models were tuned to provide some insight on how $\Sigma(r)$ influences measurements of the disk dynamics, their parameters were selected to ensure that the resulting cubes are similar enough that data quality is not a critical factor in those efforts.  These models have the same CO fluxes (7 Jy km s$^{-1}$) and sizes (280 au; defined as the radius that encircles 90\%\ of the flux), designed to match typical disks with $M_\ast \approx 1\,M_\odot$ host stars \citep{sanchis21,long22}.

\begin{figure*}[ht!]
    \centering
    \includegraphics[width=\linewidth]{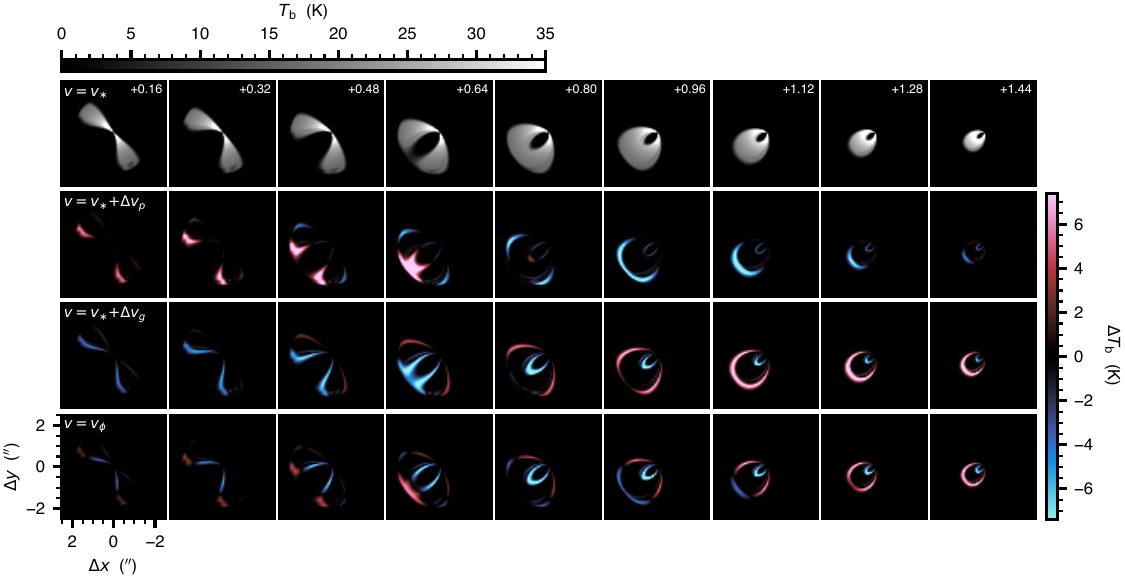}
    \caption{
    \edits{The top row shows a set of \textit{\textbf{raw}} channel maps for a variant of {\sf model B} where the velocity field is determined solely by the stellar potential: $v = v_\ast$ (Eq.~\ref{eq:vstar}).  The subsequent rows show residual maps, constructed as the differences between channel maps for other variants of the velocity field and the top row.  From top to bottom, these are for velocity field contributions that include the stellar potential plus pressure support ($v = v_\ast + \Delta v_p$, see Eq.~\ref{eq:dv}; no self-gravity), the stellar potential plus self-gravity ($v = v_\ast + \Delta v_g$, see Eq.~\ref{eq:dv}; no pressure support), and the stellar potential plus both pressure support and self-gravity ($v = v_\phi$, see Eq.~\ref{eq:bulk_motion}; the fiducal {\sf model B} case).  The annotations are as in Figure \ref{fig:chanmap_models}.     
    }\\}
    \label{fig:chanmap_contribs}
\end{figure*}

\begin{figure}[hb!]
    \centering
    \vspace{0.3cm}
    \includegraphics[width=\linewidth]{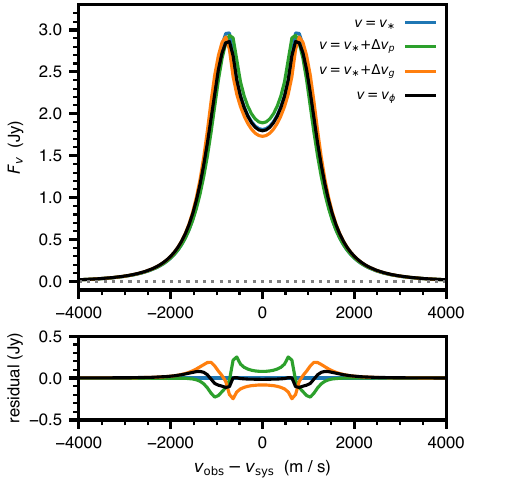}
    \caption{({\it top}) The spatially integrated spectra for {\sf model B} variants with different contributions to the velocity field (integrated from the raw cubes used to make Fig.~\ref{fig:chanmap_contribs}).  ({\it bottom}) The differences with respect to the $v_\ast$-only model (blue).  
    }  
    \label{fig:spectra_contribs}
\end{figure}

The kinematic deviations from pressure support and self-gravity impart subtle, complex changes in the emission distribution.  In Figure \ref{fig:chanmap_contribs}, these effects are decomposed for variants of raw {\sf model B} channel maps (made from distinct {\tt RADMC-3D} calculations), using the same parameters but different contributions to the net velocity field.  The top row is the reference: it includes only the stellar potential ($v = v_\ast$).  The next two rows show \edits{the residuals with respect to that reference for models with} the individual effects of pressure support ($v{=}v_\ast{+}\Delta v_p$) or self-gravity ($v{=}v_\ast{+}\Delta v_g$), respectively.  And the bottom row shows \edits{the residuals for a model with the} composite velocity field ($v = v_\phi$, defined in Eq.~\ref{eq:bulk_motion}).   

To interpret Figure \ref{fig:chanmap_contribs}, it helps to revisit the decompositions of the residual velocity profiles in Figure \ref{fig:dv_decompositions}.  The $\Delta v_p(r)$ for {\sf model B} decreases and changes sign beyond $r_t$ ($\sim$200 au).  This slowed rotation manifests as enhanced emission at larger radii near $v_{\rm sys}$, with the opposite behavior at smaller radii in the line wings.  The increasing $\Delta v_g(r)$ shifts the emission morphology to slightly faster rotation at all $r$; however, the gradient in $\Delta v_g$ generates more complex behavior than a $v_\ast$-only model with enhanced $M_\ast$, in part due to radiative transfer effects in the line wings.  The combination of these effects imparts complicated deviations for the composite field.  

Figure \ref{fig:spectra_contribs} illustrates how these same behaviors appear in a spatially-integrated sense, with spectra that demonstrate how the kinematic contributions shift around flux in $v_{\rm obs}$.  Comparing the $v{=}v_\ast{+}\Delta v_p$ and $v{=}v_\phi$ cases in Figures \ref{fig:chanmap_contribs} and \ref{fig:spectra_contribs} offers some indication of the spectral impact of varying \mdisk/$M_\ast$, from negligible ($\ll$1\%) to substantial (10\%) values.  The effects are subtle, but they are also accessible for data with sufficient resolution and sensitivity.  The next sections describe approaches for extracting such information from the data and using it to measure those quantities of interest (i.e., \mdisk).

\section{Kinematics Analysis: Concepts} \label{sec:kinematics_concepts}

\begin{figure}[ht!]
    \centering
    \includegraphics[width=\linewidth]{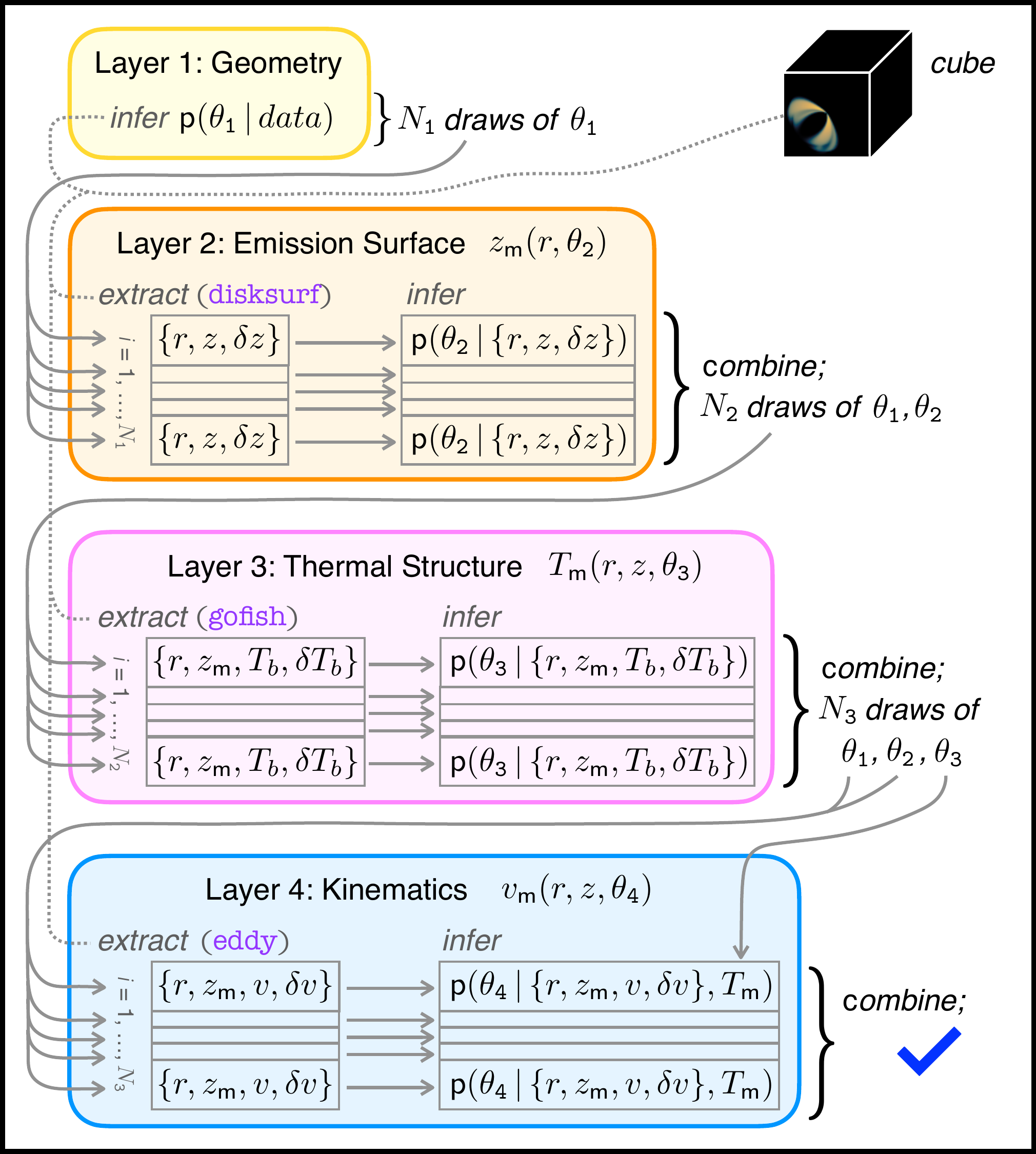}
    \caption{A flowchart of the layered analysis procedure to quantitatively interpret disk kinematics.}
    \label{fig:hierarchy}
\end{figure}

The most principled approach to measuring \mdisk\ would integrate the steps described in Sections \ref{sec:models} and \ref{sec:simobs} into a forward-modeling framework, deploying a parametric structure and radiative transfer calculations to self-consistently infer posterior distributions for the full suite of input parameters conditioned on the visibility spectra.  But that is an expensive task, since each likelihood computation is slow and the high dimensionality (there are 18 parameters in the ``simple" prescription here) of the problem requires an expansive exploration of posterior-space to quantify the globally optimized parameters.  

Fortunately, we only care here about the velocity field (though see Section \ref{sec:layer3}), and will focus on high-quality data with appropriate spatial and spectral resolution for the task.  So instead, we further developed a simplified approach \citep{veronesi21,lodato22,martire24} that forward-models a distilled measurement of $v_\phi(r)$ to constrain $\Sigma$ (and \mdisk).  This framework assumes the disk structure and kinematics are axisymmetric, and that no external contributions (e.g., cloud or envelope contamination) are present.  This section explains the process for extracting and modeling the disk kinematics.  The procedure is hierarchical, with four sequentially linked `layers' of data extraction and inference that are each outlined in more detail below.  Figure \ref{fig:hierarchy} provides a visual summary of the procedure as a guide.

\subsection{{\tt Layer\,\,1}: Geometry} \label{sec:layer1}

The geometric conversions between the observed (sky-projected) and disk spatial and spectral reference frames are critical for analyzing the spectral line emission.  The relevant coordinate transformations are specified by the spatial projection angles ($i$ and $\vartheta$), spatial translations ($\Delta x$ and $\Delta y$), and systemic velocity ($v_{\rm sys}$) in the LSRK frame.  The {\tt Layer\,\,1} task is to measure these parameters in terms of samples of the posterior probability distribution, $\mathsf{p}(\theta_{\tt 1} \,| \, data)$, where we have grouped the parameters in shorthand as $\theta_{\tt{1}}$ = [$i$, $\vartheta$, $\Delta x$, $\Delta y$, $v_{\rm sys}$] and {\it data} refers to some (here unspecified) set of measurements.  

The mechanics are left deliberately vague here, since in practice we decided to draw from {\it presumed} distributions in this layer (Sect.~\ref{sec:results}).  Often, the spatial parameters in $\theta_{\tt{1}}$ are estimated from the mm continuum \citep{dsharp2}, infrared scattered light \citep{deboer16}, or spectral line data \citep{teague21}.  A precise $v_{\rm sys}$ can be found from a straightforward measurement exploiting symmetry in the channel maps.

\subsection{{\tt Layer\,\,2}: Emission Surface} \label{sec:layer2}

The {\tt Layer\,\,2} task is to locate where the emission originates.  The radial distribution of the heights ($z$) that represent the spectral line {\it photosphere} depends on the excitation conditions and optical depths along each line-of-sight from the observer (note the implicit conditioning on $\theta_{\tt{1}}$).  Locating this emission surface requires an extraction step, where a set of surface ``coordinates" \{$r$, $z$, $\delta z$\} (with $\delta z$ representing uncertainties) are distilled from a cube, and an inference step, where those coordinates are fit with a surface model, $z_{\sf m}(r)$.

The surface coordinates were determined by locating intensity peaks in isovelocity ``loops" in channel maps deprojected into the disk-frame \citep{pinte18}, using the {\tt disksurf} package \citep{disksurf}.  The $\delta z$ were assigned to be the ratio of the beam FWHM and pixel SNR.  Coordinate extractions were limited to $160 \le |v_{\rm obs}{-}v_{\rm sys}| \le 1680$ m s$^{-1}$: further into the wings the emission is too compact, and near $v_{\rm sys}$ the deprojection is ambiguous.  An initial extraction clipped pixels outside the range $0 \le z/r \le 1$ and where $T_b \le 3.7$ K ({\tt min\char`_zr=0}, {\tt max\char`_zr=1}, {\tt min\char`_SNR=1}).  Two additional iterations were performed, using the initial surfaces to define kinematic masks that were convolved with Gaussian kernels having FWHM of 2 and 1 beams ({\tt nbeams=[2, 1]}) for clips at $T_b = 7.4$ and 11.1 K ({\tt min\char`_SNR=[2, 3]}): pixels corresponding to $z / r < 0.05$ were excluded.

A resulting set of \{$r$, $z$, $\delta z$\} was used to infer the parameters of a surface model \citep{law20},
\begin{equation}
    z_{\sf m}(r, \theta_{\tt 2}) = z_0 \bigg(\frac{r}{{\rm 150 \,\, au}}\bigg)^{\varphi} \exp{\left[-\left(\frac{r}{r_z}\right)^\psi\right]},
\end{equation}  
using $\theta_{\tt 2}$ = [$z_0$, $\varphi$, $r_z$, $\psi$, $f_{\tt surf}$] as shorthand notation for the surface parameters.  We defined a log-likelihood function $\ln{\mathcal{L}_{\tt 2}} = \ln{\mathsf{p}(\{r, z, \delta z\} \,|\, \theta_{\tt 2})}$ as
\begin{equation}
\begin{aligned}
    \ln{\mathcal{L}_{\tt 2}} &= -\frac{1}{2} \sum_{\tt i} \left[ \frac{(z_{\tt i}{-}z_{\sf m}(r_{\tt i}))^2}{s_{\tt i}^2} + \ln{2 \pi s_{\tt i}^2} \right], \\
    {\rm with} \,\, s_{\tt i}^2 &= \delta z_{\tt i}^2 + f_{\tt surf}^2 \, z_{\sf m}(r_{\tt i}^{\phantom{2}})^2. 
    \label{eq:lnlike_2}
\end{aligned}
\end{equation}
The nuisance parameter $f_{\tt surf}$ is designed to account for under-estimated scatter in $z$.  There is an implicit conditioning on $\theta_{\tt 1}$ in $\mathcal{L}_{\tt 2}$, related to the \{$r$, $z$, $\delta z$\} extraction.  

To propagate the {\tt Layer\,\,1} information into the emission surfaces, we drew $N_{\tt 1}$ random $\theta_{\tt 1}$ from the posterior samples and extracted the corresponding set of surface coordinates \{$r$, $z$, $\delta z$\} for each.  Then we sampled the (log-)posterior distribution for each draw,
\begin{equation}
    \ln{{\sf p}(\theta_{\tt 2} \,|\, \{r, z, \delta z\})} \propto \ln{\mathcal{L}_{\tt 2}} + \ln{{\sf p}(\theta_{\tt 2})}, 
    \label{eq:lnprob_2}
\end{equation}
where ${\sf p}(\theta_{\tt 2})$ is the prior, using the {\tt emcee} Monte Carlo Markov Chain sampler (\citealt{foreman-mackey13}; see Sect.~\ref{sec:results}).  The resulting $N_{\tt 1}$ sets of posterior samples were combined.  That {\it \"uber-set} of posterior samples is too large to propagate.  Instead, we drew a representative subset of $N_{\tt 2}$ values $\{\theta_{\tt 1}$, $\theta_{\tt 2}\}$ from it to use in the next layer.

\subsection{{\tt Layer\,\,3}: Thermal Structure} \label{sec:layer3}

The aim of {\tt Layer\,\,3} is to provide some constraints on the disk temperature structure, and thereby the vertical density distribution and associated kinematic contributions.  The idea is to assume the line emission is optically thick and thermalized.  In that case, the radial intensity profile is a proxy for the gas temperature profile along the emission surface.  With a judicious selection of priors, this offers some (limited) constraint on $T(r, z)$.  

The extraction step in this layer starts with computing the peak intensity (moment-8) map from the cube with the {\tt bettermoments} package \citep{bettermoments1,bettermoments2}.  Next, the {\tt gofish} package \citep{gofish} was used on that peak intensity map to extract an azimuthally-averaged brightness temperature (using the full Planck function, not the Rayleigh-Jeans limit) profile along the emission surface -- a set of \{$r$, $z_{\sf m}$, $T_b$, $\delta T_b$\} -- for each \{$\theta_{\tt 1}$, $\theta_{\tt 2}$\} draw from {\tt Layer\,\,2}.  

For the inference step, we interpreted those brightness temperature profiles with the model 
\begin{equation}
    T_{\sf m}(r, z, \theta_{\tt 3}) = T(r, z_{\sf m}),
\end{equation}
as defined in Eq.~(\ref{eq:Trz})--(\ref{eq:Tr}), with $\theta_{\tt 3}$ = [$T_{{\rm m},0}$, $T_{{\rm a},0}$, $q$, $a_z$, $w_z$, $f_{\tt temp}$].  For a single line, this is an under-constrained parameter-space.  Nevertheless, this setup is useful to maintain in the hierarchy: it allows us to meaningfully propagate that ambiguity into the inference in the final layer, even if based largely on the priors.  Moreover, in cases where observations of multiple lines are available that probe a range of emission surface heights, it can naturally incorporate that added information into improved constraints on the thermal structure.

The inference procedure is the same as for {\tt Layer\,\,2}.  We defined a Gaussian (log-)likelihood function $\ln{\mathcal{L}_{\tt 3}} = \ln{\mathsf{p}(\{r, z_{\sf m}, T_b, \delta T_b\} \,|\, \theta_{\tt 3})}$ as
\begin{eqnarray}
    \ln{\mathcal{L}_{\tt 3}} &=& -\frac{1}{2} \sum_{\tt i} \left[ \frac{(T_b(r_{\tt i}){-}T_{\sf m}(r_{\tt i}, z_{\sf m}))^2}{s_{\tt i}^2} + \ln{2 \pi s_{\tt i}^2} \right], \nonumber \\
    {\rm with} \,\, s_{\tt i}^2 &=& \delta T_b(r_{\tt i})^2 + f_{\tt temp}^2 \, T_{\sf m}(r_{\tt i}^{\phantom{2}}, z_{\sf m})^2. 
    \label{eq:lnlike_3}
\end{eqnarray}
where $f_{\tt temp}$ is a nuisance parameter designed to account for any under-estimated scatter.  Again, there is implicit conditioning on $\theta_{\tt 1}$ and $\theta_{\tt 2}$ in the data ($T_b$ profile) extraction.  For each \{$\theta_{\tt 1}$, $\theta_{\tt 2}$\} draw and associated \{$r$, $z_{\sf m}$, $T_b$, $\delta T_b$\}, the (log-)posterior distribution
\begin{equation}
    \ln{{\sf p}(\theta_{\tt 3} \,|\, \{r, z_{\sf m}, T_b, \delta T_b\})} \propto \ln{\mathcal{L}_{\tt 3}} + \ln{{\sf p}(\theta_{\tt 3})} 
\end{equation}
was sampled as before; the $N_{\tt 2}$ sets of posterior samples were combined and a representative subset of $N_{\tt 3}$ values \{$\theta_{\tt 1}$, $\theta_{\tt 2}$, $\theta_{\tt 3}$\} were then drawn to proceed.

\subsection{{\tt Layer\,\,4}: Kinematics} \label{sec:layer4}

This final layer follows the theme set in the previous ones, with an extraction and inference that is now focused on the kinematics.  The parameters of interest in {\tt Layer\,\,4} describe the contributions of the stellar and disk potentials to the gas motions -- $M_\ast$ and the prescription for $\Sigma(r)$ (and thereby \mdisk), respectively.

For each draw of \{$\theta_{\tt 1}$, $\theta_{\tt 2}$\} from {\tt Layer\,\,3} ($\theta_{\tt 3}$ does not factor in yet), we extracted a velocity profile along the emission surface using the {\tt eddy} package \citep{eddy}.  The pixel coordinates in the cube were transformed to the disk-frame along the emission surface $z_{\sf m}(r)$ and decomposed into concentric annuli.  Each annulus includes a collection of pixel spectra with different peak velocities.  The velocity $v$ in that annulus was set to the systemic value that aligns these pixel spectra so that the width of their average (stacked) spectrum is minimized ({\tt fit\char`_method=`dV'}; \citealt{teague18a}).  Each annulus is considered independently to build up a realization of the profile \{$r$, $z_{\sf m}$, $v$, $\delta v$\} for each {\tt Layer\,\,3} draw. 

Each extracted velocity profile is then interpreted in the context of the model for the velocity field,
\begin{equation}
    v_{\sf m}(r, z, \theta_{\tt 4}) = v_\phi(r, z_{\sf m}),
\end{equation}
as defined in Eq.~(\ref{eq:bulk_motion}) and the rest of Section \ref{sec:kinematics}, where $\theta_{\tt 4}$ = [$M_\ast$, \mdisk, $p$, $\gamma$, $r_t$, $f_{\tt vphi}$].  The (log-)likelihood function $\ln{\mathcal{L}_{\tt 4}} = \ln{\mathsf{p}(\{r, z_{\sf m}, v, \delta v\} \,|\, \theta_{\tt 4}})$ was defined as
\begin{eqnarray}
    \ln{\mathcal{L}_{\tt 4}} &=& -\frac{1}{2} \sum_{\tt i} \left[ \frac{(v(r_{\tt i}){-}v_{\sf m}(r_{\tt i}, z_{\sf m}))^2}{s_{\tt i}^2} + \ln{2 \pi s_{\tt i}^2} \right], \nonumber \\
    {\rm with} \,\, s_{\tt i}^2 &=& \delta v(r_{\tt i})^2 + f_{\tt vphi}^2 \, v_{\sf m}(r_{\tt i}^{\phantom{2}}, z_{\sf m})^2,
\end{eqnarray}
where $f_{\tt vphi}$ is a nuisance parameter to account for any under-estimated scatter, and there is implicit conditioning on \{$\theta_{\tt 1}$, $\theta_{\tt 2}$, $\theta_{\tt 3}$\} (the $\theta_{\tt 3}$ contribution is subtle, but it helps set the density vertical distribution and therefore the kinematic deviations associated with $\varepsilon_p$).  As in the previous layers, for each \{$\theta_{\tt 1}$, $\theta_{\tt 2}$, $\theta_{\tt 3}$\} draw and associated \{$r$, $z_{\sf m}$, $v$, $\delta v$\}, the (log-)posterior distribution
\begin{equation}
    \ln{{\sf p}(\theta_{\tt 4} \,|\, \{r, z_{\sf m}, v, \delta v\})} \propto \ln{\mathcal{L}_{\tt 4}} + \ln{{\sf p}(\theta_{\tt 4})}
\end{equation}
was sampled, and then the $N_{\tt 3}$ sets of those samples were combined into a final \"uber-set.

\section{Kinematics Analysis: Results} \label{sec:results}

Acknowledging that the hierarchical inference strategy outlined above seems (perhaps unpleasantly) complicated, this section guides the reader through various assumptions to demonstrate how the different aspects of the procedure work and how certain factors contribute to the precision and accuracy of the key parameters.

\subsection{\edits{Spatial Resolution} Bias} \label{sec:res_bias}

We start with some direct `fidelity' tests of the modeling infrastructure, vastly oversimplifying the parameter-space by assigning $\delta$-function priors at the input values (denoted with asterisk superscripts): $\mathsf{p}(\theta) = \delta(\theta-\theta^\ast)$.  Such priors apply to all parameters except the subset of $\theta_{\tt 4}$ comprising [$M_\ast$, \mdisk] (and $f_{\tt vphi}$).  This means we effectively skipped over {\tt Layers\,\,1-3} in the hierarchy.  The primary goal is to illustrate that bias enters the problem even in such an (unrealistic) omniscient state.  

First, we need to establish the true emission surface parameters $\theta_{\tt 2}^\ast$: unlike all the other parameters, $\theta_{\tt 2}$ are not explicitly specified.  We used the {\tt tausurf} mode of raytracing in {\tt RADMC-3D} to generate cubes of the disk-frame coordinates ($r$, $\phi$, $z$) where the line-of-sight optical depth $\tau_{\rm los} = 2/3$ (i.e., using the Eddington approximation to define the photosphere).  We associated the surface with the coordinates that correspond to the frequency of peak emission in each pixel.  Since the line emission is so optically thick, these surfaces trace the tops of the CO-rich layers ($z_{\sf m}(\theta_{\tt 2}^\ast) \approx z_{\rm crit}$ in Eq.~\ref{eq:scrit}), as was assumed above (e.g., in making Figure \ref{fig:dv_decompositions}).  Table \ref{table:surfaces} compiles the results for each fiducial model.     

\begin{deluxetable}{c | c c c c}[!t]
\tablecaption{$\tau_{\rm los} = 2/3$ Surface Parameters, $\theta_{\tt 2}^\ast$ \label{table:surfaces}}
\tablehead{
\colhead{} & \colhead{$z_0$ (au)} & \colhead{$\varphi$} & \colhead{$r_z$ (au)} & \colhead{$\psi$}
}
\startdata
\textsf{model A}\phantom{sprea} & \phantom{a}43\phantom{a} & \phantom{a}1.07\phantom{a} & \phantom{a}323\phantom{a} & \phantom{a}3.99\phantom{a}  \\
\textsf{model B}\phantom{sprea} & 43 & 1.06 & 309 & 6.51  \\
\textsf{model C}\phantom{sprea} & 42 & 1.05 & 313 & 10.61 
\enddata
\end{deluxetable}

\begin{figure}[t!]
    \includegraphics[width=\linewidth]{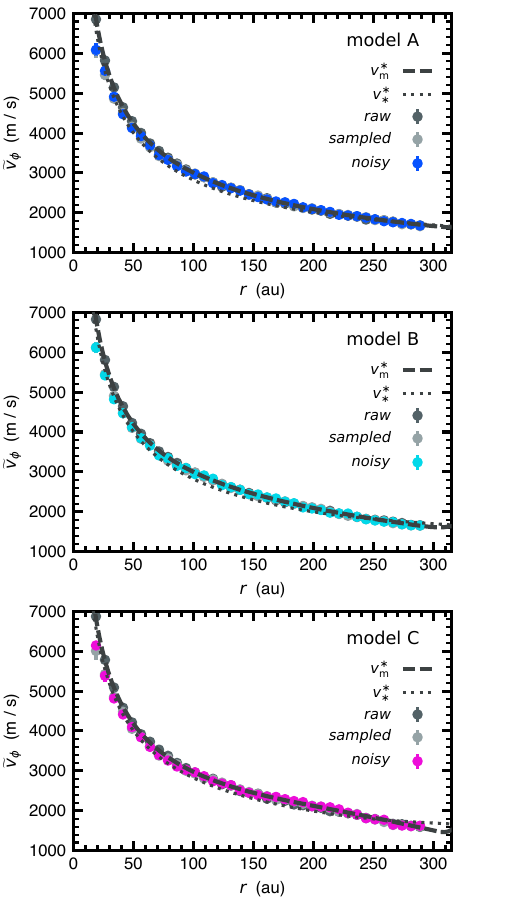}
    \caption{The extracted $\widetilde{v}_\phi(r)$ profiles from the raw, sampled, and noisy versions of each model cube, compared with the true model $v_{\sf m}^\ast(r)$ and the ``background" field set by the stellar potential, $v_\ast^\ast(r)$ along the emission surfaces (Table \ref{table:surfaces}).  Notice how the extracted $\widetilde{v}_\phi(r)$ profiles from cubes with limited spatial resolution (sampled and noisy) are systematically biased low (by $\lesssim 10$\%) in the inner disk ($r \lesssim 100$ au).}
    \label{fig:vphi_51}
\end{figure}

With the priors set (to the values in Tables \ref{table:fid_params} and \ref{table:surfaces}), we performed the {\tt Layer\,\,4} extractions of velocity profile data for the raw, sampled, and noisy cubes.  We adopted radial bins spaced by 0\farcs05 (roughly half the synthesized beam FWHM) and extending out to where the line intensity is below the noise floor.  Figure \ref{fig:vphi_51} shows the extracted velocities for each cube type and fiducial model, compared with the associated true profiles ($v_{\sf m}^\ast$) and the contributions from the stellar potential alone ($v_\ast^\ast$).

\begin{figure}[t!]
    \includegraphics[width=\linewidth]{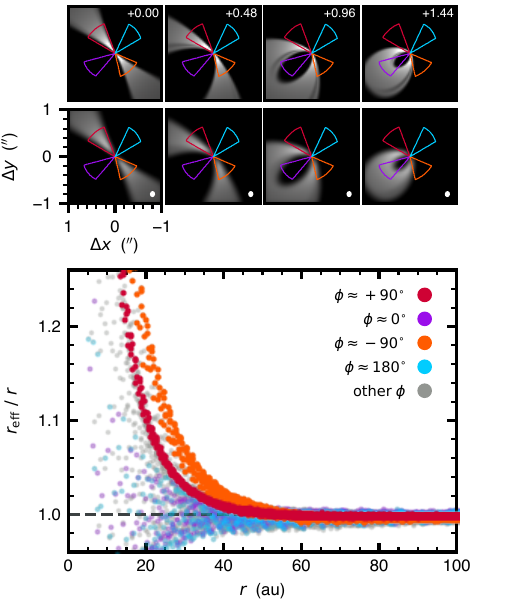}
    \caption{\edits{The top rows show a small subset of channel maps for the {\sf model B} raw cube (top) and the raw cube convolved with a Gaussian kernel matched to the synthesized beam (marked in the lower right corners), appropriate for the sampled (and noisy) cube.  These show every sixth 80 m s$^{-1}$ channel to span a broader (projected) spectral range, with LSRK velocities (in km s$^{-1}$) marked in the upper right corners of the top row.  The bottom plot shows the ratio $r_{\rm eff} / r$ for all pixels, color-coded by some representative azimuth ranges (as marked in the channel maps).  This highlights that the beam-averaged intensity distributions in the inner disk generally behave like the pixels from larger radii, and thereby will have effective velocities that are biased low.  There is a complex azimuthal behavior, however, which means that a direct analytic correction is a challenge.  Instead, we adopted an empirical correction process that works sufficiently well.}}
    \label{fig:beamsmear}
\end{figure}

Overall, the extracted velocities $\widetilde{v}_{\phi}(r)$ match the true profile well, recovering the behavior of the subtle residuals induced by pressure support and self-gravity across most of the disk.  But at smaller radii ($\lesssim 100$ au), the extractions from the sampled and noisy cubes are biased low.  This is an artifact of the spatial resolution: it is \edits{a consequence of the beam-averaging of gradients in the projected intensity distribution (e.g., see \citealt{keppler19} or \citealt{boehler21}).  Those intensity gradients are most pronounced in the channel maps at small radii (and compact line wings).  Spatial filtering (limited $u$, $v$-coverage) has negligible contribution here: the bias for the sampled cube is nearly identical to that measured from convolving the raw cube with a Gaussian beam.  However, sparse spatial frequency sampling might produce an analogous bias in less ideal datasets.} 

\edits{Figure \ref{fig:beamsmear} illustrates the origins of this resolution bias.  In terms of extracting $\widetilde{v}_\phi(r)$ from a cube, the {\it effective radius} ($r_{\rm eff}$) of each pixel can be approximated with the average of the true $r$ in each pixel weighted by the intensity distribution.  For the raw cube (top row), the pixels are independent and there is no bias: $r_{\rm eff} = r$, the true deprojected radius in the disk-frame along the emission surface $z_{\sf m}^\ast$ (Table \ref{table:surfaces}).  But when we convolve the raw cube with a Gaussian beam (matched to the synthesized beam for the sampled cube; next row in Fig.~\ref{fig:beamsmear}), the beam-averaged intensities from neighboring pixels factor into the averages.  At larger $r$, the convolved intensity gradients on beam scales are modest, so $r_{\rm eff} \approx r$.  But at smaller $r$, there are steeper intensity gradients present on sub-beam scales that generally push $r_{\rm eff} \gtrsim r$.  The bottom panel of Figure \ref{fig:beamsmear} demonstrates this effect.  While there is a complicated azimuthal dependence, the net impact is clear: limited spatial resolution makes the properties of the emission in these regions behave like they originate at larger radii, which (given the velocity gradient) translates into a bias to diminished $\widetilde{v}_{\phi}(r)$.}   

It is hard to predict where and how this effect becomes problematic in general, since it depends on a non-trivial combination of the disk velocity field, viewing geometry, and resolution.  But this bias is a significant concern at radii well beyond the resolution scale, extending here to $\gtrsim 4\times$ the synthesized beam FWHM.  \edits{Moreover, the bias is independent of the $\widetilde{v}_\phi(r)$ extraction algorithm: it is an intrinsic property of the data, and will be an issue for any approach that does not forward-model the effects of limited spatial resolution.}  As we demonstrate below, this bias needs to be corrected upon extraction to avoid propagating it into the {\tt Layer\,\,4} inference.  

Fortunately, a straightforward (if crude) debiasing approach works well.  Assuming the bias is produced solely by these beam-smearing effects, a radial profile of correction factors $\mathcal{C}_v(r)$ can be determined empirically by comparing velocity extractions from {\it similar} mock data with known velocity fields and intensity gradients.  For each dataset, we used {\tt csalt} (Andrews et al., {\it in preparation}) to generate a mock cube with the same pixels and channels and the median extrinsic properties determined in {\tt Layer\,\,1}.  The mock cube assumed the emission originates in a thin layer defined by the median $z_{\sf m}(r)$ inferred in {\tt Layer\,\,2}, with a power-law brightness temperature profile set to the median $T_{\sf m}(r)$ along that surface inferred in {\tt Layer\,\,3}.  We assumed linewidths were thermal, and optical depths were high (1000 at 150 au) and decreased inversely with $r$ (although those two assumptions are less important here).  The mock velocity field was set to $v^{\,\sf mock}_\phi = v^{\,\sf mock}_\ast(M_\ast)$ (Eq.~\ref{eq:vstar}) along the emission surface, with $M_\ast$ assigned based on a least-squares fit to the median $\widetilde{v}_\phi(r)$ extracted from the dataset.  The mock cube was then convolved with a Gaussian kernel matched to the synthesized beam.  We then followed Section \ref{sec:layer4} and extracted $N_{\tt 3}$ realizations of $\widetilde{v}^{\,\sf mock}_\phi(r)$ from the mock cube using the \{$\theta_{\tt 1}$, $\theta_{\tt 2}$, $\theta_{\tt 3}$\} draws determined from the data (the noisy cubes) in {\tt Layer\,\,4}.

\begin{figure}[t!]
    \includegraphics[width=\linewidth]{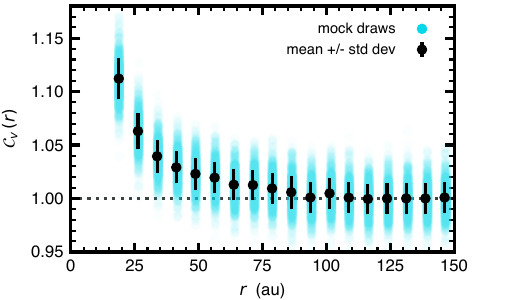}
    \caption{The velocity correction factors measured empirically for {\sf model B} using $N_{\tt 3} = 2,000$ extractions from a (convolved) mock cube.  The averages over all mocks in each radial bin were adopted to adjust for the resolution smearing effect that biases the extracted velocities at smaller $r$.}
    \label{fig:vphi_corr}
\end{figure}

\begin{figure*}[t!]
    \begin{minipage}[c]{0.5\linewidth}
        \centering
        \includegraphics{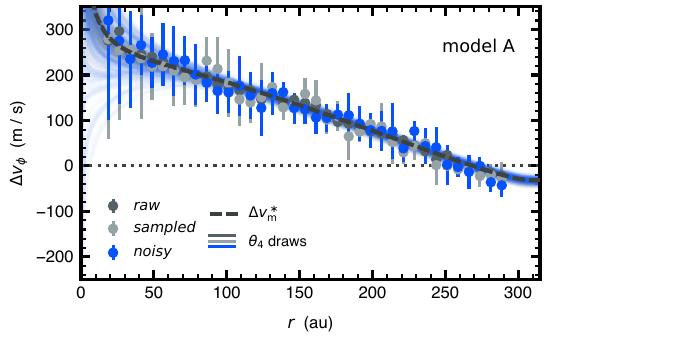}
        \includegraphics{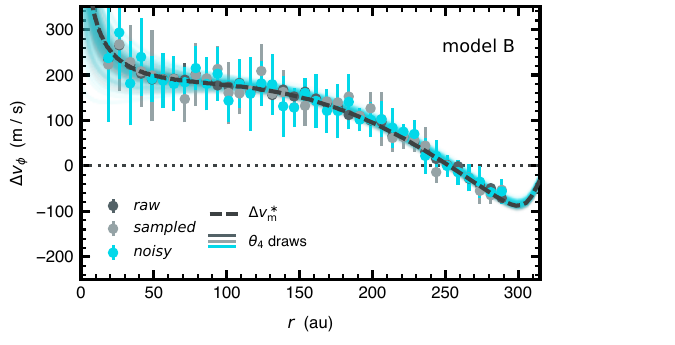}
        \includegraphics{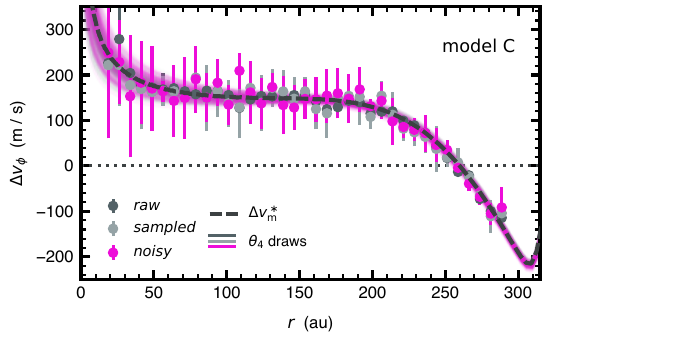}        
    \end{minipage}\hfill
    \begin{minipage}[c]{0.5\linewidth}
        \centering
        \includegraphics{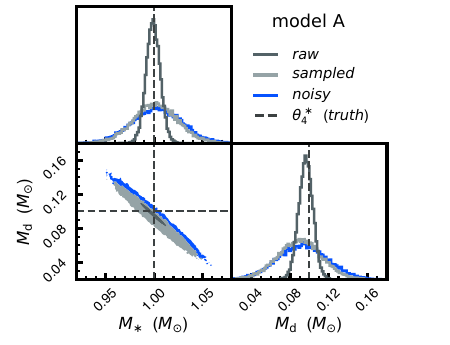}
        \includegraphics{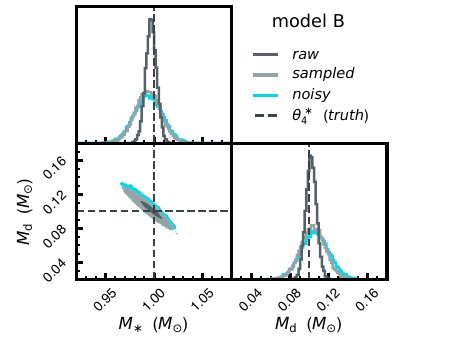}
        \includegraphics{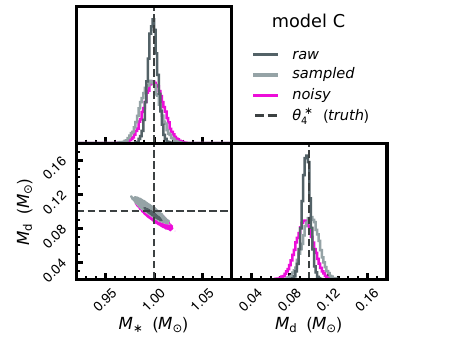}
    \end{minipage}
    \caption{({\it left}) The extracted (and bias-corrected) residual velocity profiles, referenced to the true stellar potential profile, $\Delta v_\phi = v_\phi - v_\ast^\ast$, from the raw, sampled, and noisy cubes for each fiducial model.  These are compared with the true profile along the emission surface ($\Delta v_{\sf m}^\ast$) and profiles constructed from 100 random draws of the posteriors inferred assuming $\delta$-function priors for $\theta_{\tt 1}$, $\theta_{\tt 2}$, $\theta_{\tt 3}$, and the $\Sigma(r)$ parameters set at their true values, and the $M_\ast$ and \mdisk\ priors in Eq.~(\ref{eq:prior4_simple}).  ({\it right}) The marginalized posteriors, with the true values marked with dashed lines (contours mark the 2$\sigma$ confidence intervals). \\ }
    \label{fig:dvphi_51}
\end{figure*}

We defined the correction factors as 
\begin{equation}
    \mathcal{C}_v(r) = \frac{v^{\,\sf mock}_\ast(r)}{\widetilde{v}^{\,\sf mock}_{\phi}(r)},
\end{equation}
the ratio of the input velocities to the (beam-convolved, biased) extracted values for the mock cube.  The $\mathcal{C}_v(r)$ measured from the mocks for {\sf model B} are shown in Figure \ref{fig:vphi_corr}.  The $\mathcal{C}_v(r)$ are indistinguishable for the three fiducial models and relatively insensitive to uncertainties in the input parameters (within $<$1\%\ standard deviations), confirming the basic assumption that beam-smearing of the projected spatio-kinematic field (scaling with the same, well-constrained factor $\sqrt{M_\ast} \sin{i}$) is the origin of the bias.  From this point, we consider the {\tt Layer\,\,4} ``data" as the {\it bias-corrected velocities} (based on the correction factors derived from the mocks), defined as $v_\phi(r) = \widetilde{v}_\phi(r) \overline{\mathcal{C}_v}(r)$ (the overline denoting the average over the mocks; e.g., the black points in Figure \ref{fig:vphi_corr}).

\begin{figure*}[t]
    \begin{minipage}[c]{0.5\linewidth}
        \centering
        \includegraphics{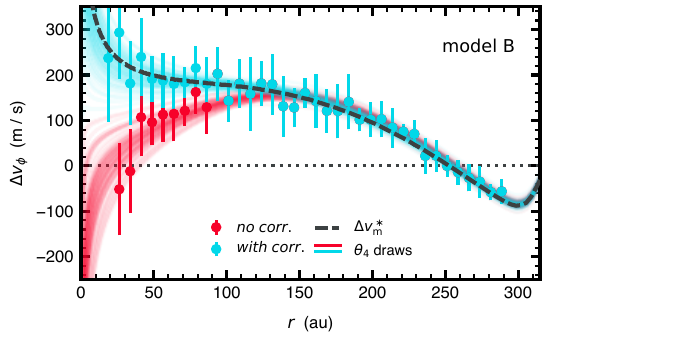}
    \end{minipage}\hfill
    \begin{minipage}[c]{0.5\linewidth}
        \centering
        \includegraphics{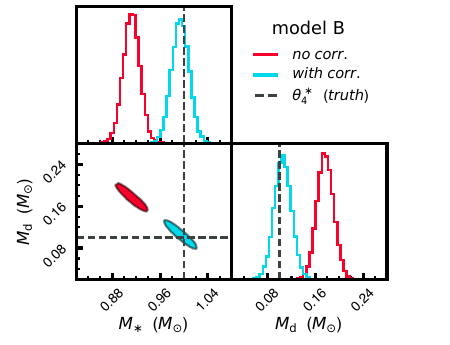}
    \end{minipage}
    \caption{A comparison of variations on the extracted (residual) velocities for the {\sf model B} noisy cube and their influence on the inferences of $M_\ast$ and \mdisk, following the same prescription and notations as in Figure \ref{fig:dvphi_51}.  These include the {\it uncorrected} $\widetilde{v}_\phi(r)$ (red) and the {\it corrected} $v_\phi(r)$ (blue: the same as in Figure \ref{fig:dvphi_51}).  The limited spatial resolution of the data biases the extracted velocities low, out to a radius much larger than the synthesized beam FWHM.  Those artificially diminished velocities (red) produce substantial biases in the physical parameters of interest.  An approximate empirical correction, derived presuming that the projected velocity field is biased by beam convolution, substantially mitigates those inaccuracies (blue). \\}
    \label{fig:vphi_bias}
\end{figure*}

With those considerations, we performed the {\tt Layer\,\,4} inference in the special case where $\theta_{\tt 1}$, $\theta_{\tt 2}$, $\theta_{\tt 3}$, and the $\Sigma(r)$ parameters $p$, $\gamma$, $r_t$ are fixed at their true values.  We adopted uniform priors for the limited subset of $\theta_{\tt 4}$, 
\begin{equation}
    {\sf p}(\theta_{\tt 4}) = 
    \begin{cases}
        {\sf p}(M_\ast) & \hspace{-0.85cm} \sim \mathcal{U}(0, 5\,\,M_\odot) \\
        {\sf p}(M_{\rm d}) & \hspace{-0.85cm} \sim \mathcal{U}(0, 1\,\,M_\odot) \\
        {\sf p}(\ln{f_{\tt vphi}}) & \hspace{-0.30cm} \sim \mathcal{U}(-10, 2).
    \end{cases}
    \label{eq:prior4_simple}
\end{equation}
We configured a standard {\tt emcee} setup to have 64 walkers, initialized by random draws from the priors.  Here, we sampled the posteriors for 10,000 steps and assessed convergence based on the autocorrelation times $\langle \tau \rangle$ (in this case $\sim$50 steps).  We discarded the initial $20 \langle \tau \rangle$ steps (as the burn-in phase) and thinned by retaining every $0.5 \langle \tau \rangle$ step from the remaining samples for each walker.  Figure \ref{fig:dvphi_51} shows the results of those inferences, comparing the {\it residual} velocity profiles (defined here by reference to the true stellar potential profile, $\Delta v_\phi = v_\phi - v_\ast^\ast$, solely for ease of visualization; note the actual quantity used in the inference calculations is $v_\phi$) with posterior draws for the raw, sampled, and noisy cubes of each model.  The marginalized posterior distributions for $M_\ast$ and \mdisk\ are presented to the right in each row.  The estimated uncertainties from the profile extractions ($\delta v$) are appropriate: no evidence for inflated scatter was found ($f_{\tt vphi} < 0.009$ at 99\%\ confidence).     

The inferences for all cube types overlap well, with peaks that recover the true $M_\ast$ well within 1\%\ and only modest ($\lesssim$10\%) biases on \mdisk, along a clear degeneracy (anti-correlation).  This is the accuracy limit induced by the procedure outlined in Section \ref{sec:layer4}, at least in this specific case of rigid model restrictions.  Note that the inferred precision improves for steeper $\Sigma(r)$ tapers, because it is easier to differentiate pressure support and self-gravity when the kinematic impact of the taper is confined to a narrow radial range (e.g., $\sim$230--280 au for {\sf model C}, compared to $\sim$75--280 au for {\sf model A}).

Before relaxing the modeling restrictions imposed by the strict priors, it is instructive to more directly illustrate the effects of the resolution bias noted above.  Figure \ref{fig:vphi_bias} compares the results of inferences for the {\sf model B} noisy cube using either the {\it uncorrected} $\widetilde{v}_\phi(r)$ (red) or the {\it corrected} $v_\phi(r)$, as in Figure \ref{fig:dvphi_51} (blue).  The resolution bias imparts significant biases on $M_\ast$ and especially \mdisk, pushing the posteriors along their covariance to lower $M_\ast$ ($\sim$10\%) and higher \mdisk\ ($\sim$170\%).  The stellar potential ($v_\ast$ term in Eq.~\ref{eq:bulk_motion}) is more important in the inner disk, so the resolution bias to lower $v_\phi$ at small $r$ pushes the $M_\ast$ inference lower.  To compensate for the reduced velocities at larger $r$, where self-gravity is important, the inferred \mdisk\ is pushed higher.  The amount of bias depends on the $\Sigma(r)$ taper: it is more problematic for lower $\gamma$, where the taper affects a larger radial range ({\sf model A}).  Ignoring a correction for this resolution bias will deliver misleading \mdisk\ measurements.

\subsection{Geometry and Surface Effects} \label{sec:geosurf}

Next, we explore the downstream effects that are associated with precision and accuracy in the geometry ({\tt Layer\,\,1}) and emission surfaces ({\tt Layer\,\,2}).  Like Section \ref{sec:res_bias}, we set $\delta$-function priors at the true values for $\theta_{\tt 3}$ (skipped {\tt Layer\,\,3}) and a subset of $\theta_{\tt 4}$ ($p$, $\gamma$, and $r_t$).  In {\tt Layer\,\,1}, we drew $N_{\tt 1} = 500$ samples from $\theta_{\tt 1}$ ``posteriors" that we {\it assigned} -- in practice, this means drawing from adopted priors ${\sf p}(\theta_{\tt 1})$.  We assumed that these priors were independent and normally distributed $\sim \mathcal{N}(\mu, \sigma)$, with means set at their true values (Table \ref{table:fid_params}) and standard deviations comparable to those inferred in the literature for similar targets \citep[e.g.,][]{huang18,andrews21},
\begin{equation}
    {\sf p}(\theta_{\tt 1}) = 
    \begin{cases}
        {\sf p}(\Delta x) & \hspace{-0.35cm} \sim \mathcal{N}(0.00, 0.01\arcsec) \\
        {\sf p}(\Delta y) & \hspace{-0.35cm} \sim \mathcal{N}(0.00, 0.01\arcsec) \\
        {\sf p}(i) & \hspace{-0.70cm} \sim \mathcal{N}(30, 0.5\degr) \\
        {\sf p}(\vartheta) & \hspace{-0.65cm} \sim \mathcal{N}(130, 0.5\degr) \\
        {\sf p}(v_{\rm sys}) & \hspace{-0.25cm} \sim \mathcal{N}(0, 4\,\,{\rm m \,\, s}^{-1}).
    \end{cases}
    \label{eq:prior1}
\end{equation}
These $\theta_{\tt 1}$ draws were used to extract $N_{\tt 1}$ sets of surface coordinates \{$r$, $z$, $\delta z$\} in {\tt Layer\,\,2}.  

We then inferred $\theta_{\tt 2}$ posterior samples for each draw, adopting uniform priors for the surface parameters,
\begin{equation}
    {\sf p}(\theta_{\tt 2}) = 
    \begin{cases}
    {\sf p}(z_0) & \hspace{-1.00cm} \sim \mathcal{U}(0, 150 \,\, {\rm au}) \\
    {\sf p}(\varphi) & \hspace{-1.05cm} \sim \mathcal{U}(0, 5) \\
    {\sf p}(r_z) & \hspace{-1.03cm} \sim \mathcal{U}(0, 450 \,\, {\rm au}) \\
    {\sf p}(\psi) & \hspace{-1.05cm} \sim \mathcal{U}(1, 15), \\
    {\sf p}(\ln{f_{\tt surf}}) & \hspace{-0.25cm} \sim \mathcal{U}(-10, 2). 
    \end{cases}
    \label{eq:prior2}
\end{equation}
The $\theta_{\tt 2}$ posteriors were sampled with {\tt emcee}, using 64 walkers for 10,000 steps.  The mean autocorrelation time for those samples was $\langle \tau \rangle \approx 60$ steps.  The mean value of the nuisance parameter $f_{\tt surf} \approx 1$, indicating the surface uncertainties ($\delta z$) are under-estimated by a factor $\sim$5--10 (as expected, given the large scatter found for any individual extracted set of surface coordinates).

\begin{figure}[ht!]
    \includegraphics[width=\linewidth]{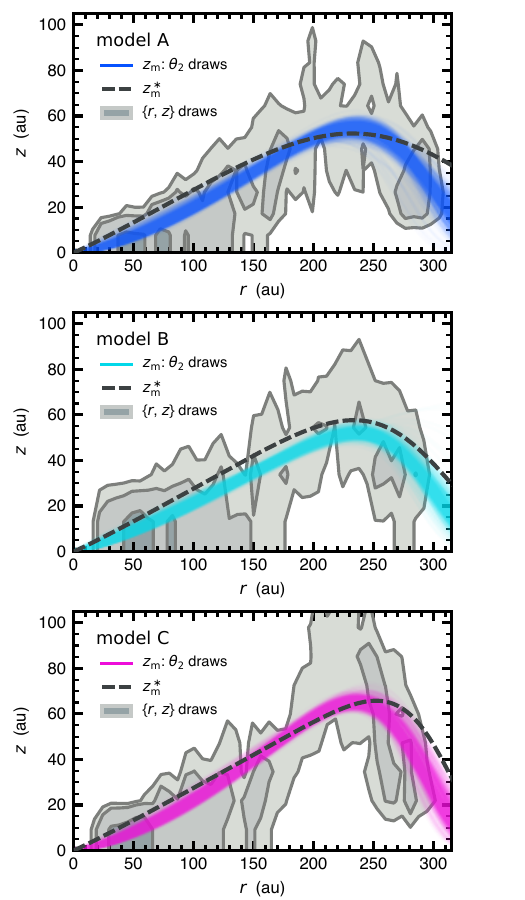}
    \caption{Histogram summaries of the extracted surface coordinates \{$r$, $z$\} for the noisy cubes of each fiducial model (contours mark 15, 50, and 85\%\ of the coordinates) from the $N_{\tt 1}$ realizations in {\tt Layer\,\,2}, compared with the $z_{\sf m}$ profiles from 1,000 draws of the $\theta_{\tt 2}$ posteriors and the true $z_{\sf m}^\ast$ profiles. 
    }
    \label{fig:surf_52}
\end{figure}

Figure \ref{fig:surf_52} summarizes the emission surface inferences in {\tt Layer\,\,2}.  The $\theta_{\tt 2}$ posteriors produce $z_{\sf m}(r)$ that mimic the shape of the true surface, but are systematically low (a mean offset $\Delta z / r \approx -0.05$).  That bias is caused by two factors.  First is a radiative transfer effect (diagnosed with the raw cubes\edits{, in some experiments with the back surface removed in the {\tt RADMC-3D} calculations}), where {\tt disksurf} finds shifted emission lobe centroids because of the asymmetric intensity ``shoulders" produced over a range of line-of-sight depths into the disk.  These are exacerbated for higher inclinations, thicker CO-rich layers, and steeper vertical temperature gradients.  Second is a spatial resolution effect, where the emission lobe peaks become difficult to separate at smaller $r$.  

\begin{figure*}[t!]
    \begin{minipage}[c]{0.5\linewidth}
        \centering
        \includegraphics{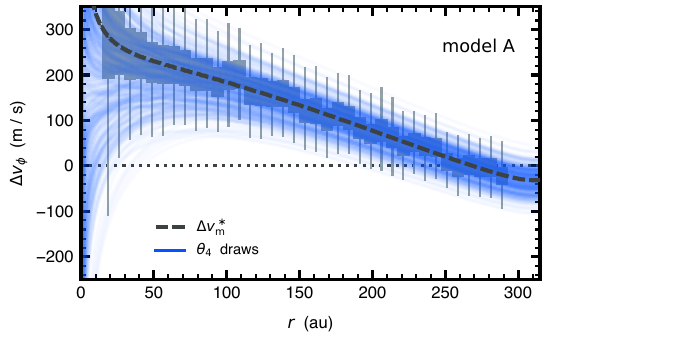}
        \includegraphics{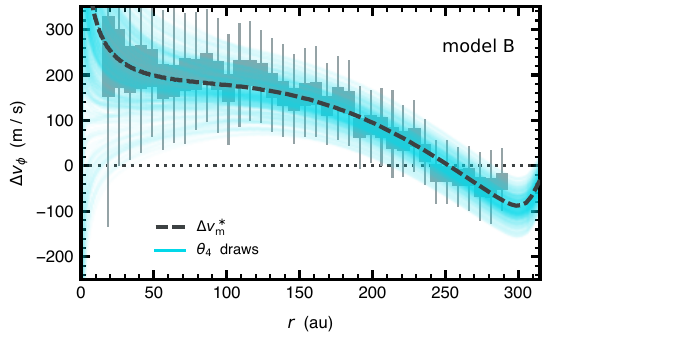}
        \includegraphics{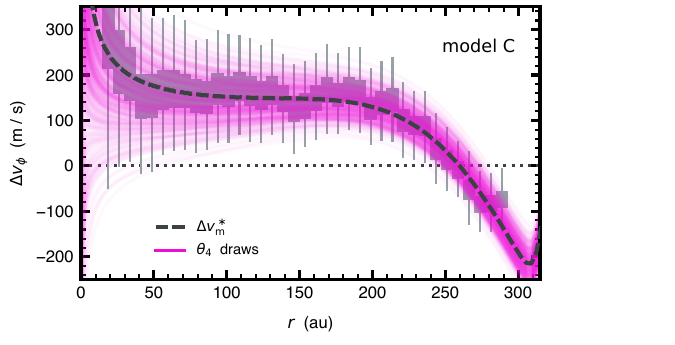}        
    \end{minipage}\hfill
    \begin{minipage}[c]{0.5\linewidth}
        \centering
        \includegraphics{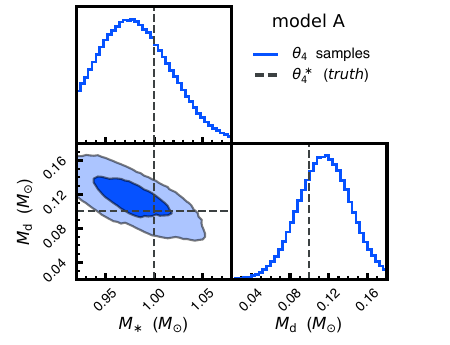}
        \includegraphics{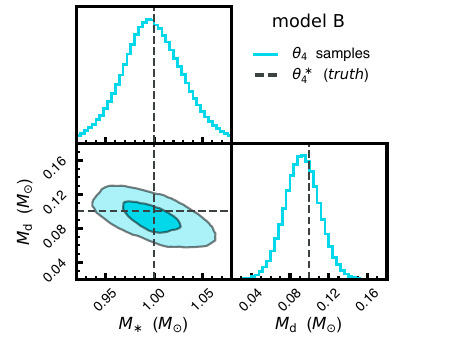}
        \includegraphics{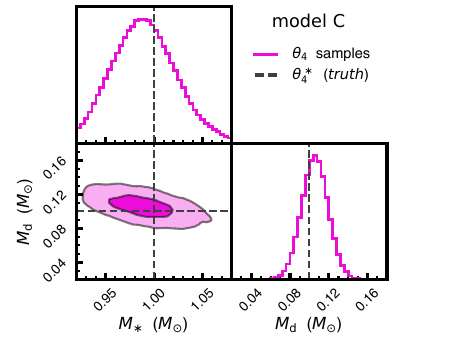}
    \end{minipage}
    \caption{As in Figure \ref{fig:dvphi_51}, a comparison of the residual velocity profiles ({\it left}) and marginalized posteriors for the restricted model inferences ({\it right}), now accounting for the ambiguities in the geometry and emission surfaces.  The box-whisker plots show the distributions of the velocity extractions: the boxes denote the interquartile range (50\%) and the whiskers (error bars) span 99\%\ of the extractions.  The covariance corner plots show contours at the 1 and 2$\sigma$ confidence intervals. \\}
    \label{fig:vphi_52}
\end{figure*}

A subset of $N_{\tt 2} = 1,000$ draws from the \{$\theta_{\tt 1}$, $\theta_{\tt 2}$\} posterior samples collected in {\tt Layer\,\,2} were propagated to {\tt Layer\,\,4} to extract realizations of $\{r, v, \delta v\}$.  The bias-corrected $v_\phi(r)$ from this ensemble spans a range with a standard deviation that is $\sim$30\%\ larger than the formal uncertainties in any individual draw (35--65 m s$^{-1}$ compared to 25--50 m s$^{-1}$).  That additional scatter is dominated by ambiguities in the geometric parameters; uncertainties in the emission surfaces are negligible.  The mean extracted $v_\phi(r)$ is similar to the true input ($v_{\sf m}^\ast$), suggesting the geometry and surface ambiguities do not impart a strong bias in the velocity extractions.  Using the same {\tt emcee} setup as in Section \ref{sec:res_bias}, we sampled the (subset of) $\theta_{\tt 4}$ posteriors for each draw and combined the (burned-in, thinned) results (the mean autocorrelation time was $\langle \tau \rangle \approx 50$ steps).  As before, there is no need to account for additional scatter in the velocity profile extractions ($f_{\tt vphi} < 0.01$ at 99\%\ confidence).

Figure \ref{fig:vphi_52} summarizes the (subset of) $\theta_{\tt 4}$ posteriors and corresponding $v_\phi(r)$ in this case, where the $\theta_{\tt 3}$ and \{$p$, $\gamma$, $r_t$\} parameters in $\theta_{\tt 4}$ are held fixed by $\delta$-function priors at the true values.  The peaks of the marginal posteriors for $M_\ast$ and \mdisk\ do not shift much compared to Section \ref{sec:res_bias} (Figure \ref{fig:dvphi_51}); the small biases noted are largely tied to the rigidity of the model given the adopted priors.  This confirms that the surface bias noted in Figure \ref{fig:surf_52} does not propagate into a significant $v_\phi$ bias.  That is not surprising, since the vertical variation of the velocity field in the disk models used to generate the data are modest (Figure \ref{fig:dv}).  The posterior widths for $M_\ast$ and \mdisk\ broaden (roughly by a factor of two) when the geometry and surface ambiguities are considered, tracing the increased range of the extracted $v_\phi(r)$ profiles due primarily to the ambiguities in $\theta_{\tt 1}$.  But we do not read too much into these posterior morphologies because the stringent priors limit the flexibility in the inference.

\subsection{The Complete, Layered Inference} \label{sec:final_inference}

Having now explored the isolated consequences of the geometric and emission surface parameters, we relax the stringent constraints on the parameters describing the physical conditions (i.e., $\theta_{\tt 3}$ and the rest of $\theta_{\tt 4}$) and run the full hierarchical workflow (Section \ref{sec:kinematics_concepts}; Figure \ref{fig:hierarchy}).  Following the steps outlined above, we started from the {\tt Layer\,\,2} outputs and extracted $N_{\tt 2} = 1,000$ realizations of the $T_b(r)$ profile corresponding to the subset of draws from the \{$\theta_{\tt 1}$, $\theta_{\tt 2}$\} posterior samples.  

Next, we assigned a joint conditional $\theta_{\tt 3}$ prior, 
\begin{equation}
    {\sf p}(\theta_{\tt 3}) = 
    \begin{cases}
            {\sf p}(T_{{\rm m}, 0}) & \hspace{-0.65cm} \sim \mathcal{N}(45, 8 \,\, {\rm K}) \\
            {\sf p}(T_{{\rm a}, 0}) & \hspace{-0.70cm} \sim \mathcal{U}(0, 400 \,\, {\rm K}) \\
            {\sf p}(q) & \hspace{-1.15cm} \sim \mathcal{N}(0.5, 0.1) \\
            {\sf p}(a_z) & \hspace{-1.00cm} \sim \mathtt{Exp}(10) \\
            {\sf p}(w_z) & \hspace{-0.95cm} \sim \mathtt{Exp}(20) \\
            {\sf p}(\ln{f_{\tt temp}}) & \hspace{-0.25cm} \sim \mathcal{U}(-10, 2)
    \end{cases}
    \label{eq:prior3}
\end{equation}
if $T_{{\rm m}, 0} \le T_{{\rm a}, 0}$, or ${\sf p}(\theta_{\tt 3}) = 0$ otherwise.  The $\sim \mathtt{Exp}(\lambda)$ expression for $a_z$ and $w_z$ in Eq.~(\ref{eq:prior3}) denotes an exponential prior ${\sf p}(x) = \lambda e^{-\lambda x}$.  We adopted this form to deprioritize large values, where the prescription is problematic (the $\tanh$ in Eq.~\ref{eq:Tz} does not have finite support).

The prior for $T_{{\rm m}, 0}$ was assigned based on a simplified prescription for a passive irradiated disk,
\begin{equation}
    T_{{\rm m}, 0} \approx \left[ \frac{\eta L_\ast}{8 \pi \sigma_\textsc{sb} r_0^2} \right]^{1/4},
    \label{eq:tmid}
\end{equation}
where $\eta$ is the flaring angle of the disk surface, $L_\ast$ is the stellar luminosity, and $\sigma_\textsc{sb}$ is the Stefan-Boltzmann constant \citep{chiang97,dullemond01}.  Adopting a log-normal $L_\ast$ prior with a typical uncertainty \citep{alcala17} that peaks at the \citet{baraffe15} model value for a 1 $M_\odot$ star at 2 Myr, ${\sf p}(\log_{10}{(L_\ast/L_\odot)}) \sim \mathcal{N}(0.08, 0.15)$, and a liberal uniform prior for the flaring angle, ${\sf p}(\eta) \sim \mathcal{U}(0.01, 0.1)$, the prescription in Eq.~(\ref{eq:tmid}) suggests a $T_{{\rm m}, 0}$ prior reasonably approximated with Eq.~(\ref{eq:prior3}).  Note the slight bias compared to the true value ($T_{\rm m, 0}^\ast = 40$ K; Table \ref{table:fid_params}).    

\begin{figure}[t!]
    \includegraphics[width=\linewidth]{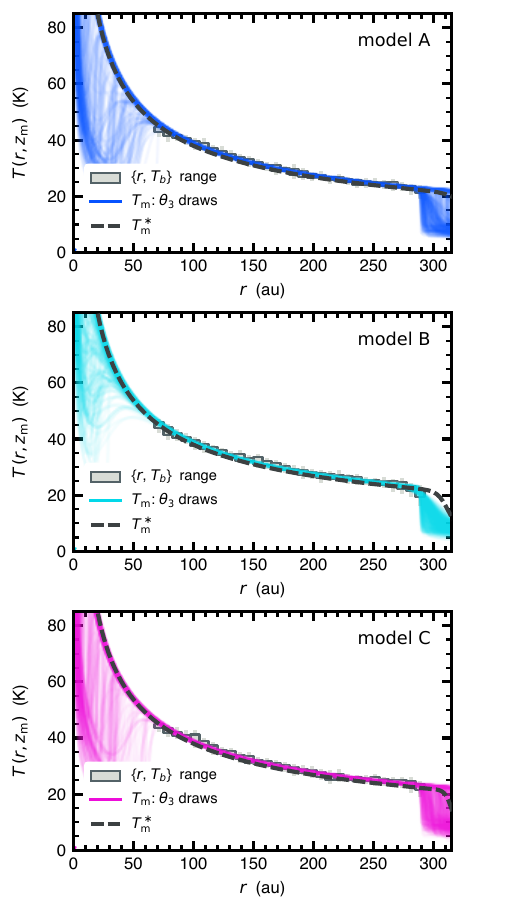}
    \caption{Summaries of the extracted $T_b(r)$ profiles from the $N_{\tt 3}$ realizations in {\tt Layer\,\,3}, compared with the reconstructed $T_{\sf m}(r, z_{\sf m})$ from 500 draws of the $\theta_{\tt 3}$ posteriors and the true temperature profile along the emission surface $T_{\sf m}^\ast(r, z_{\sf m}^\ast)$.}
    \label{fig:temp_53}
\end{figure}

We then proceeded with the inference step in {\tt Layer\,\,3} and sampled the $\theta_{\tt 3}$ posteriors for 64 walkers and 10,000 steps (mean autocorrelation time $\langle \tau \rangle \approx 140$ steps).  The additional scatter term here was small, with $f_{\tt temp} < 0.01$ (99\%\ confidence).  Figure \ref{fig:temp_53} summarizes the results: the $\theta_{\tt 3}$ posteriors are somewhat biased off the true values, as expected for the very limited constraints the $T_b(r)$ profile along a single emission surface offers.  The adopted priors are perhaps unfairly stringent, since they dominate the inferences for $T_{{\rm m}, 0}$.  However, if multiple spectral line datasets can be combined, the {\tt Layer\,\,3} inferences can be less reliant on such priors.

\begin{figure*}[ht!]
    \centering
    \includegraphics[width=\linewidth]{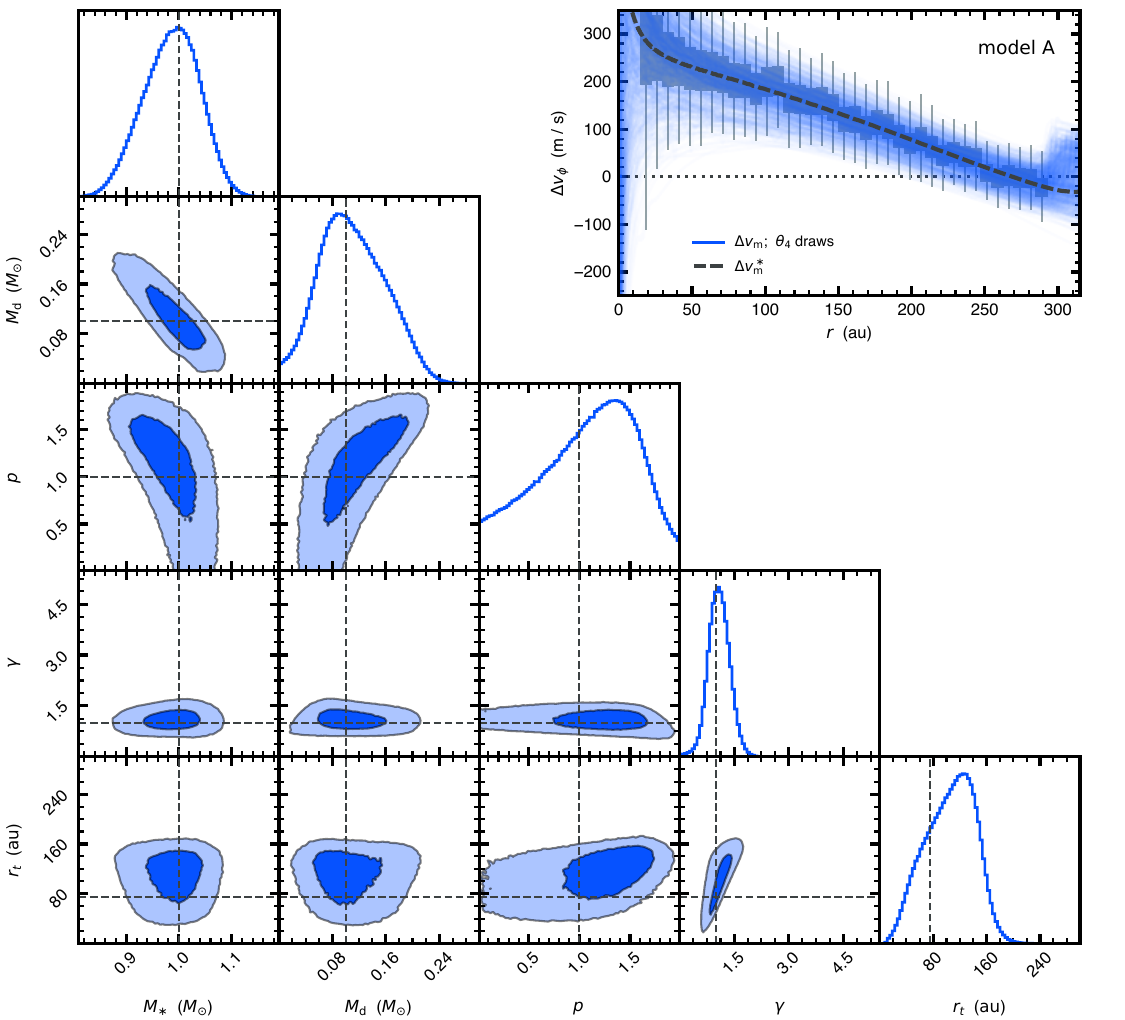}
    \caption{The marginalized $\theta_{\tt 4}$ posteriors for {\sf model A}, compared with $\theta_{\tt 4}^\ast$: contours mark the 1 and 2$\sigma$ confidence intervals.  The panel in the upper right shows a binned summary of the extracted (residual) velocity profile data from the $N_{\tt 3}$ realizations in {\tt Layer 4} (as in Figure \ref{fig:vphi_52}), compared with the reconstructed $\Delta v_{\sf m}$ from 1,000 draws of the $\theta_{\tt 4}$ posteriors and the true $\Delta v_{\sf m}^\ast$. \\}
    \label{fig:layer4_53A}
\end{figure*}

Next, we drew a subset of $N_{\tt 3}\,{=}\,2,000$ realizations of \{$\theta_{\tt 1}$, $\theta_{\tt 2}$, $\theta_{\tt 3}$\} from the {\tt Layer\,\,3} \"uber-set of posteriors and propagated those to {\tt Layer\,\,4} to extract $N_{\tt 3}$ velocity profile datasets \{$r$, $v$, $\delta v$\}.  Each of those (bias-corrected) $v_\phi(r)$ realizations was then used to infer posterior samples for the full complement of $\theta_{\tt 4}$, again using {\tt emcee} and 64 walkers for 25,000 steps.  In addition to the priors specified in Eq.~(\ref{eq:prior4_simple}), we adopted  
\begin{equation}
    {\sf p}(\theta_{\tt 4}) = 
    \begin{cases}
        {\sf p}(p) & \hspace{-0.37cm} \sim \mathcal{U}(0, 2) \\
        {\sf p}(\gamma) & \hspace{-0.35cm} \sim \mathcal{U}(0, 6) \\
        {\sf p}(r_t) & \hspace{-0.25cm} \sim 1 - \mathtt{Logistic}(300 \,\, {\rm au}, 0.1),
    \end{cases}
    \label{eq:prior4}
\end{equation}
where the $\mathtt{Logistic}(x_0, b)$ distribution is 
\begin{equation}
    {\sf p}(x) = \left(1 + e^{-b (x - x_0)} \right)^{-1}.
\end{equation}
In this case, ${\sf p}(r_t)$ acts like a soft upper bound (in practice, the adopted $x_0$ simply means we assume the $\Sigma$ taper happens at a radius where we still detect CO emission, as suggested by the $z_{\sf m}(r)$ and $\Delta v_\phi(r)$ behaviors).  The Eq.~(\ref{eq:prior4}) priors are designed to be effectively agnostic.  We also included a composite prior that ensures the solutions are gravitationally stable,
\begin{equation}
    {\sf p}({\tt min} \, Q) = \mathtt{Logistic}(2, 20),  
    \label{eq:Qprior}
\end{equation}
with ${\tt min} \, Q$ the minimum Toomre parameter (Eq.~\ref{eq:toomre}).  This prior is designed so that instability accrues a steep penalty, but stable models (${\tt min} \, Q \gtrsim 2$) are equally probable.  We consider this a well-motivated prior for massive, axisymmetric disks, which helps avoid artificially broadening the \mdisk\ posteriors to non-physical values.  If \mdisk\ is lower, it is essentially negligible.

\begin{figure*}[ht!]
    \centering
    \includegraphics[width=\linewidth]{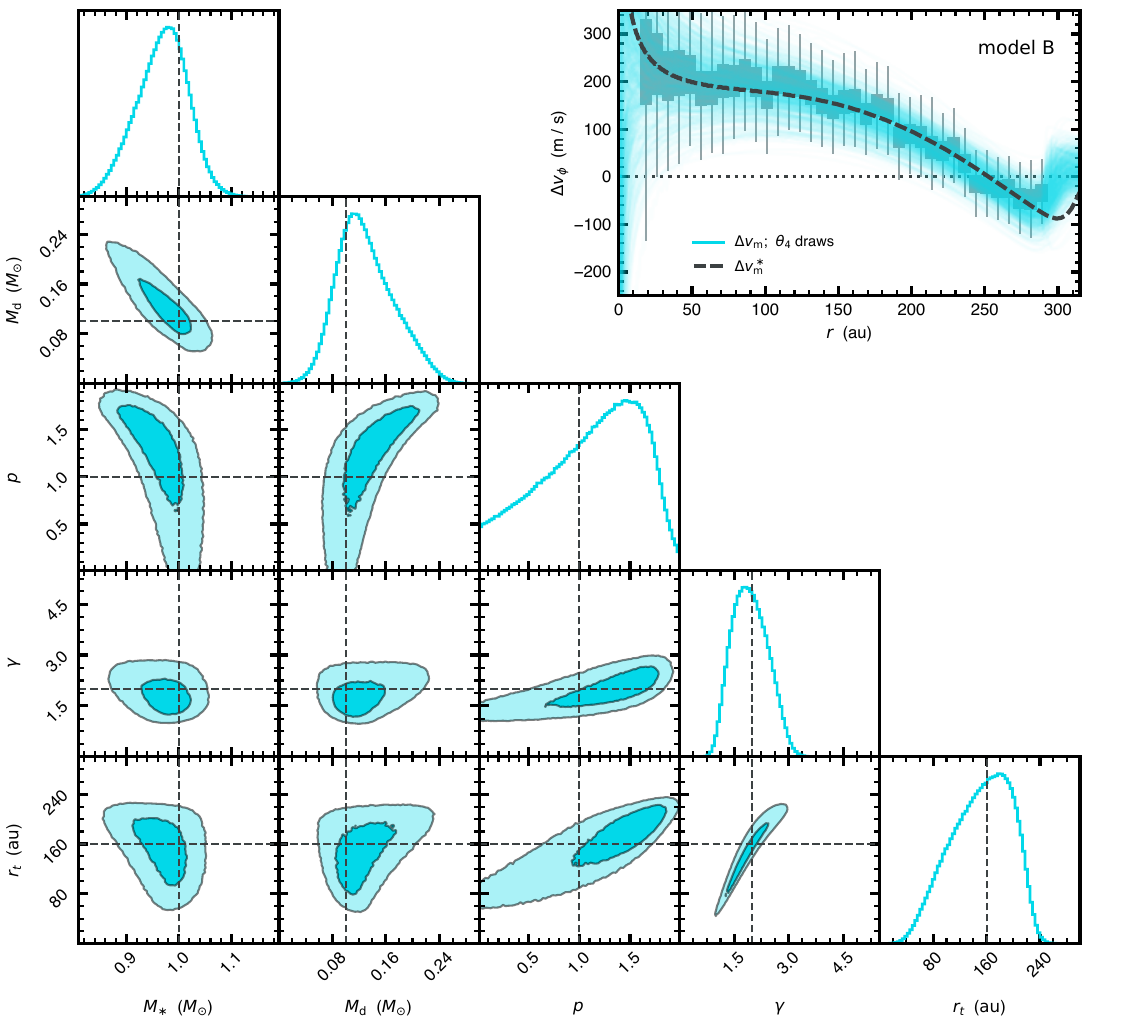}
    \caption{The same information and annotations as in Figure \ref{fig:layer4_53A}, in this case for {\sf model B}.   ~~~~~~~~~~~~~~~~~~~~~~~~~~~~~~~~~~~~~~~~~~~~~~ \\}
    \label{fig:layer4_53B}
\end{figure*}

\begin{figure*}[ht!]
    \centering
    \includegraphics[width=\linewidth]{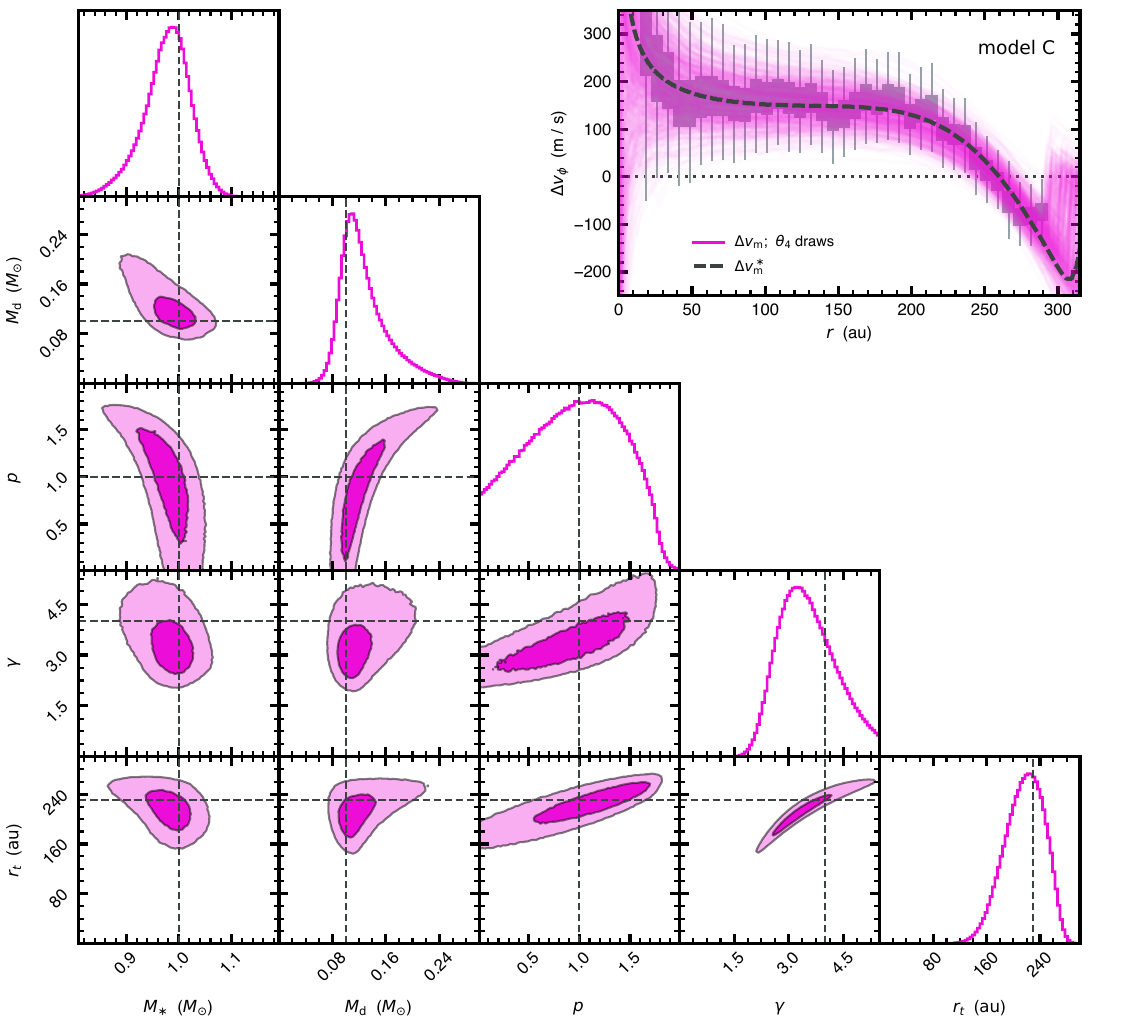}
    \caption{The same information and annotations as in Figure \ref{fig:layer4_53A}, in this case for {\sf model C}. ~~~~~~~~~~~~~~~~~~~~~~~~~~~~~~~~~~~~~~~~~~~~~~ \\}
    \label{fig:layer4_53C}
\end{figure*}

The {\tt Layer\,\,4} inference results are presented in Figures \ref{fig:layer4_53A}--\ref{fig:layer4_53C}, which show marginalized projections for pairs of $\theta_{\tt 4}$ posterior samples along with comparisons of the extracted (residual) velocities and $\Delta v_\phi(r)$ reconstructed from random posterior draws for each fiducial model.  The mean autocorrelation time was $\langle \tau \rangle \approx 200$ steps.

The inferred $M_\ast$--\mdisk\ covariance behavior is similar to what was noted for the more restricted inferences in Sections \ref{sec:res_bias} and \ref{sec:geosurf}.  However, the flexibility afforded by relaxing the priors on the $\theta_{\tt 4}$ parameters that describe $\Sigma(r)$ imparts a more general complexity on the results.  In all cases, the marginalized $M_\ast$ and \mdisk\ posteriors are consistent with the true values: biases for these parameters are $\lesssim$\,2\% and $\lesssim$\,20\%, respectively.

The precision of the \mdisk\ measurement depends on the $\Sigma(r)$ taper.  Broader posteriors (less precision) are found for more gradual tapers (lower $\gamma$): the 90\%\ confidence intervals for \mdisk\ span roughly $\pm$0.08 $M_\odot$ for {\sf model A} ($\gamma = 1$), $\pm$0.06 $M_\odot$ for {\sf model B} ($\gamma = 2$), and $\pm$0.04 $M_\odot$ for {\sf model C} ($\gamma = 4$).  As we noted above, the underlying difference is the radial range over which the density taper matters.  For steep tapers like {\sf model C}, that range is limited and the kinematic signature is pronounced: it is therefore easier to constrain the detailed shape of $\Sigma(r)$ from the dynamical contributions of both pressure support and self-gravity.  But for gradual tapers like {\sf model A}, much more of the disk is affected; this accentuates the kinematic degeneracy between $\varepsilon_p$ and $\varepsilon_g$.

\begin{figure}[ht!]
    \centering
    \includegraphics[width=\linewidth]{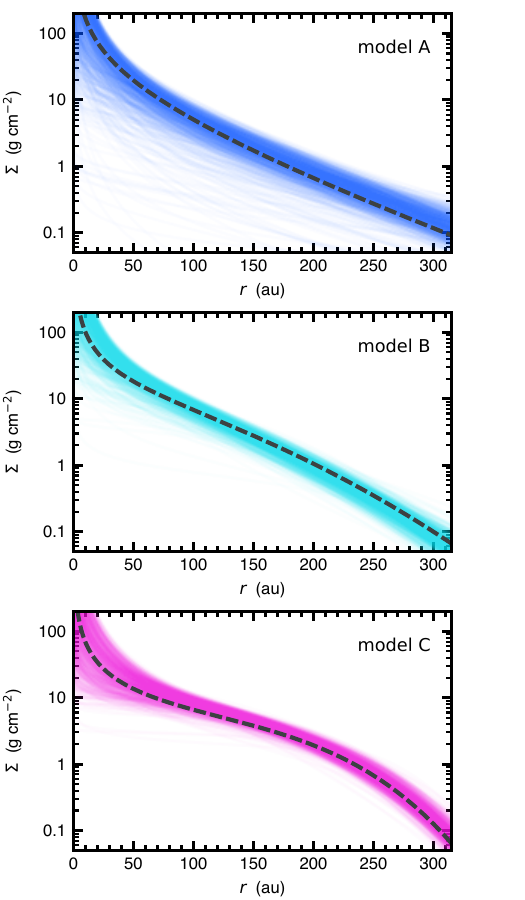}
    \caption{The reconstructed surface density profiles using 1,000 draws from the $\theta_{\tt 4}$ posteriors for each fiducial model, compared with the true profiles (dashed).}
    \label{fig:sigmas}
\end{figure}

Despite these details, the overall quality of the surface density profile retrievals for these fiducial models -- using solely the kinematics information in spectral line measurements -- is remarkably good.  Figure \ref{fig:sigmas} shows $\Sigma(r)$ profiles reconstructed from the $\theta_{\tt 4}$ posterior draws in each case, compared with the true (input) profiles.  The bulk of the posterior draws find $\Sigma(r)$ profiles within a factor of two of the true profiles over essentially the full range of radii that produces CO line emission.

\section{Discussion} \label{sec:discussion}

\subsection{Applicability and Limitations} \label{sec:main_disc}

Considering the analyses presented in Section \ref{sec:results} holistically, a few important lessons can be distilled from this methodological study.  First, it is essential to account for the resolution bias that impacts velocity profile extractions at small radii to retrieve an accurate assessment of the mass distribution (Sect.~\ref{sec:res_bias}): ignoring this effect injects a factor of $\sim$two bias (high) in the density normalization (\mdisk), which will unfortunately compound with other factors to undermine the effort.  Second, assuming unbiased estimates of the geometric parameters, accounting for a reasonable level of ambiguity in the emission surfaces does not bias the kinematics: the typical levels of geometric uncertainty introduce only modestly degraded precision for the density inferences ($\sim$50\%\ in \mdisk; Sect.~\ref{sec:geosurf}).  Third, intrinsic degeneracies regarding the shape of $\Sigma(r)$ are the determining factor for the quality of \mdisk\ inferences (Sect.~\ref{sec:final_inference}), with less precision for gradual outer disk tapers (i.e., lower $\gamma$).

\begin{figure}[ht!]
    \centering
    \includegraphics[width=\linewidth]{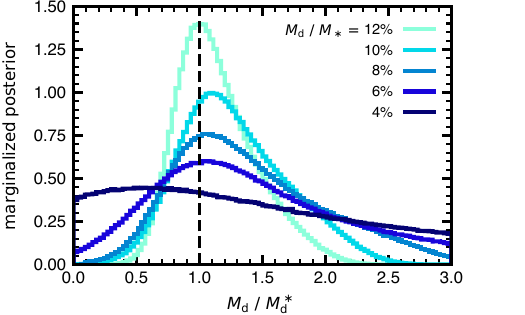}
    \caption{The marginalized \mdisk\ posteriors inferred for a sequence of {\sf model B} variants with different $M_{\rm d} / M_\ast$ values (note that the 10\%\ case is the fiducial {\sf model B} case), following the procedure outlined in Section \ref{sec:kinematics_concepts}.  Each distribution is normalized by the true \mdisk, to highlight the level of posterior broadening as the mass decreases.  The transition from a firm measurement (with good fidelity) to an upper limit occurs around $M_{\rm d} / M_\ast \approx 5\%$ in this example.  }
    \label{fig:mratio}
\end{figure}

Using the layered analysis strategy developed for this task (Sect.~\ref{sec:kinematics_concepts}), we found that we could infer $\Sigma(r)$ with remarkably good fidelity (i.e., within a factor of two) for data of appropriate quality and reasonable assumptions if the disk is an appreciable fraction of the total mass, $M_{\rm d} / M_\ast \approx 10$\%\ (Fig.~\ref{fig:sigmas}).  As that ratio decreases, the kinematic signature from $\varepsilon_g$ decreases and the quality of the \mdisk\ retrieval is diminished.  To explore that behavior, we generated a suite of analogous synthetic data for variants of {\sf model B} with $M_{\rm d} = 0.04$, 0.06, 0.08, and 0.12 $M_\odot$.  Aside from scaling the column density threshold for the upper boundary of the CO-rich layer ($\sigma_{\rm crit})$ to compensate for the different \mdisk\ and maintain a fixed emission size, the other input parameters were fixed to the values in Table \ref{table:fid_params}.  We followed the same full inference procedure as above for each of these variants.  The marginalized \mdisk\ posteriors for each (including the fiducial {\sf model B}) are shown together in Figure \ref{fig:mratio}, with each distribution normalized relative to the true mass.

The distributions in Figure \ref{fig:mratio} illustrate the anticipated behavior: the \mdisk\ posteriors broaden as the intrinsic mass ratio decreases.  If all else is equal, and these models are representative, this experiment suggests that the approach outlined here could deliver good quality \mdisk\ inferences down to $M_{\rm d} / M_\ast \approx 5\%$ without any additional modifications.  At face value, this suggests the approach may only be relevant at the high end of the \mdisk\ distribution.  But recall that spectral line {\it intensity}-based \mdisk\ estimates at such high values struggle with accuracy, due to the high optical depths: in that sense, the kinematics approach is a valuable complement.

There are opportunities to improve this \mdisk\ sensitivity range by combining observations of the kinematics from multiple spectral lines.  The intrinsic degeneracies in the problem are diminished with increasing precision in the extracted $v_\phi(r)$ information.  Even if it proves difficult or impractical to seek such improvements for any given spectral line, the framework described above (Sect.~\ref{sec:kinematics_concepts}) can accommodate multiple lines that, when considered together, improve the inference quality.  In practice, the multi-line approach would proceed the same way in {\tt Layer\,\,1} (since the disk has only one relevant set of geometric parameters), and then independently for each line in {\tt Layer\,\,2} and for the extraction components of {\tt Layers\,\,3} and {\tt 4}.  But the inference components of these latter two layers would be conducted jointly for the set of lines (with simple composite posteriors).

In the simplest scenario where statistical noise dominates (rather than systematics; see below), combining $N$ lines with similar $v_\phi(r)$ quality could yield $\sqrt{N}$ improvement in \mdisk\ precision (e.g., a $M_{\rm d} / M_\ast \approx 3$--4\% measurement for two tracers).  A reasonable option might be to use multiple CO transitions (e.g., CO $J$=2$-$1 and $J$=3$-$2) and/or another species with bright and optically thick lines (e.g., HCO$^+$).  This might {\it underestimate} the improvement in the case where multiple lines probe the kinematics at sufficiently different vertical surfaces, since two-dimensional information better disentangles the contributions of pressure support and self-gravity.  This is more of a challenge, but could be realized using, e.g., lines from CO isotopologues.  Realistically, such lines will have less precise $v_\phi(r)$ (they are intrinsically fainter) over a more limited radial range.  Nevertheless, the returns on folding in constraints with such complementary information, even if the quality is lower, would  certainly benefit the $\Sigma(r)$ inferences.

\subsection{Caveats and Potential} \label{sec:caveats}

While it is difficult to properly generalize the applicability of the results for a few fiducial models to the high-dimensional parameter space (geometric, physical, observational) that may be relevant for real measurements, there are some intrinsic aspects of the problem that can guide future expectations and strategies.  

From an observational perspective, while both sensitivity and spectral resolution are relevant for precision, it is especially important to {\it spatially} resolve the emission well.  Perhaps it is obvious from the discussion of the resolution bias (Sect.~\ref{sec:res_bias}), but proper sampling of $v_\phi(r)$ gradients is essential for an accurate inference of $\Sigma(r)$.  As a crude reference metric, we suggest seeking $\gtrsim$\,20 FWHM resolution elements across the emission distribution diameter.  For this reason, it will be difficult to measure a robust $\Sigma(r)$ with this technique for compact disks (emission line radii $\lesssim$\,1\arcsec) with ALMA.  

We argued in Section \ref{sec:geosurf} that the intrinsic ambiguities in our knowledge of the extrinsic geometric parameters of a disk target ($\theta_{\tt 1}$) make a substantial contribution to the scatter in the extracted $v_\phi(r)$, and therefore into the precision of the $\Sigma(r)$ inferences.  It does not seem likely that we could improve measurements of the disk center or sky-projection angles much beyond what is assumed here before other systematic factors (e.g., asymmetries) dominate.  However, the modeling procedure does depend on the {\it value} of the inclination angle (and not just the associated precision).  For moderate $i$ ($\sim$20--50\degr), we can do a reasonable job of inferring the emission surfaces and extracting good-quality $v_\phi(r)$, as described here.  At more face-on orientations, the surfaces become more ambiguous \citep[e.g.,][]{teague22}, and therefore the accuracy of the $\Sigma(r)$ inferences depend more on the priors.  At higher $i$, the $v_\phi(r)$ extractions might suffer confusion associated with the three-dimensional emission geometry (e.g., contamination from the backside emission surface).  The tradeoffs associated with the viewing geometry are complex, but will be essential to explore as the tools associated with these quantitative dynamical inferences are developed further.    

Aside from acquiring suitable data, there are some areas for immediate progress that can improve $\Sigma(r)$ inferences in this analysis framework.  For example, these measurements depend on a robust constraint of the two-dimensional thermal structure determined by combining optically thick tracers that sample different vertical layers \citep{law21,paneque-carreno23}.  But in practice, we usually reconstruct a temperature distribution from sparsely sampled (in $z$) information.  Efforts to design a better-informed, empirical prescription for  $T(r, z)$ using more tracers would retire a lingering risk of biased interpretations of the kinematics.  There are also potential improvements available from fine-tuning the algorithms used in extracting the emission surfaces and velocity profiles; for example, by selecting optimizers or prescriptions that are better aligned with realistic emission distributions (i.e., informed by radiative transfer simulations).  If such adjustments reduce bias or improve precision -- particularly in $v_\phi(r)$ -- we would see strong returns on the quality of $\Sigma(r)$ inferences.  

\edits{The over-arching aspiration for treating such {\it solvable} issues is to position the problem so that the systematics are dominated by {\it physical} characteristics; i.e., a mis-specification of the correct model, not some technical issue in the practical aspects of the measurements.}  Such mis-specification systematics could arise from complications to the analysis if various assumptions are incorrect.  While many options are possible, most of the problems can be traced back to over-simplified prescriptions for the physical conditions and/or dynamics.  Such incomplete parameterizations can bias the kinematic inferences in a way that is difficult to foresee.  While this is a generic concern for most astronomical inference problems, there is evidence for specific examples that could prove to be subtle challenges for future work.

An obvious example is the prevalence of substructures, thought to trace localized gas pressure deviations \citep[e.g.,][]{brogan15,andrews16,dsharp1,dsharp2,long18}.  If such features are unresolved and axisymmetric, the accompanying kinematic deviations ($\varepsilon_p$) are likely negligible; if they are resolved, they should be incorporated into the model (by modifying the $\Sigma$ and $T$ prescriptions accordingly).  But if any substructures are non-axisymmetric (e.g., spirals, vortices), the azimuthal averaging used in this analysis framework {\it will} generate artificial $v_\phi(r)$ structures that can bias the \mdisk\ inference: an alternative approach is required.  Another, more subtle, example is the contribution of magnetic pressure.  Some simulations predict that it could dominate the thermal pressure at $z / r \gtrsim 0.3$ \citep{hirose11,wang19}, and therefore influence the kinematics (through $\varepsilon_p$) traced by spectral lines emitting from such heights.  

In terms of the dynamics more specifically, there are various mechanisms that could introduce radial (e.g., accretion flows; \citealt{rosenfeld14,sperez15}) or vertical (e.g., winds or outflows; \citealt{teague18a,galloway-sprietsma23}) contributions to the gas motions.  These were neglected here, under the assumption that they can be accurately disentangled from $v_\phi(r)$.  However, this may not be the case in general, particularly for disks with complicated geometric projections (e.g., warps; \citealt{rosenfeld12a,casassus18}).  While the spectral line kinematics offer unique opportunities to probe these effects, it remains to be seen what level of bias such systematics can have on dynamical \mdisk\ estimates.

\subsection{Demonstration with ALMA Data: MWC 480} \label{sec:mwc480}

\begin{figure}[ht!]
    \centering
    \includegraphics[width=\linewidth]{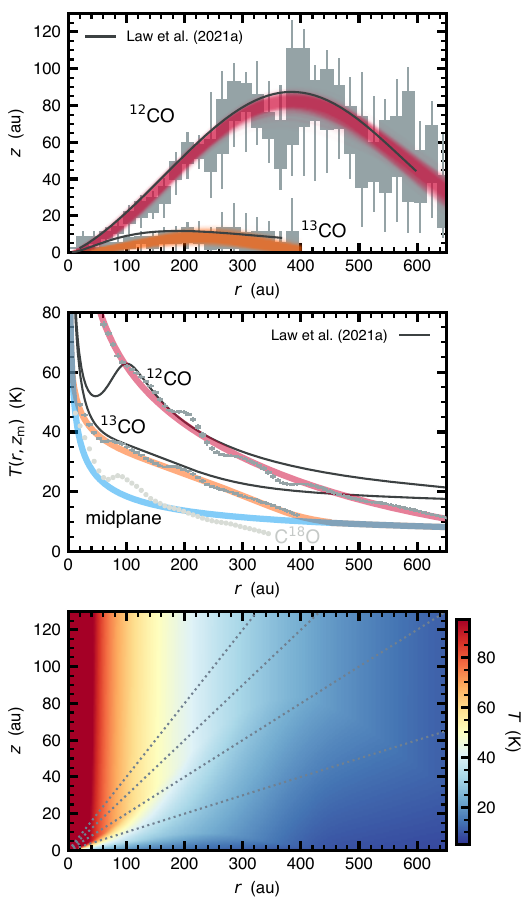}
    \caption{({\it top}) The extracted (gray box-whisker plots) and inferred (colored curves) emission surfaces for $^{12}$CO and $^{13}$CO $J$=2$-$1 in the MWC 480 disk.  Black curves mark the surfaces derived from the same data by \citet{law21}.  ({\it middle}) The $T_b(r)$ distributions (gray box-whisker plots; dispersions are small) extracted along those surfaces, compared with the adopted gas temperature models for those surfaces (red and orange curves) and the midplane (blue curve).  The corresponding \citet{law21} models are again shown as black curves.  While not used directly here, the extracted $T_b(r)$ distributions for the MAPS C$^{18}$O $J$=2$-$1 data (assuming it originates in the midplane) are also shown for reference (light gray circles).  ({\it bottom}) The median temperature distribution from the reconstruction adopted based on interpolation of the information in the top two panels, with the same scale and annotations as in Figure \ref{fig:Trz}.}
    \label{fig:MWC480_layers123}
\end{figure}

Having further developed and quantitatively validated the analysis procedure outlined in this article in a `controlled' environment, it makes sense to deploy it on real observations as a test case where the true disk properties are not known a priori.  For this, we selected the ALMA observations from the MAPS project \citep{maps1} of the MWC 480 disk.  This target has a similar geometry and CO emission surface to the fiducial models explored above \citep{law21}, appears to be axisymmetric \citep{teague21}, and does not suffer foreground or environmental contamination (unlike, for example, AS 209 or GM Aur).  Moreover, the MAPS data and the MWC 480 disk have some qualities that can help us illustrate some potential future development of the standard approach outlined in this article.  We considered the cubes for the $^{12}$CO and $^{13}$CO $J$=2$-$1 transitions provided by MAPS, created with FWHM synthesized beam dimensions $0\farcs17\times0\farcs12$ (PA = 6\degr) and a channel spacing of 200 m s$^{-1}$ (these are the {\tt robust} = $+0.5$, primary beam-corrected cubes).

The geometric parameters used in {\tt Layer\,\,1} were determined from the MAPS 225 GHz continuum visibilities, following the approach of \citet{andrews21} and presented in Appendix \ref{app:mwc480_layer1}.  We adopted $\Delta x = 5\pm7$ mas, $\Delta y = 1\pm7$ mas, $i = -36\fdg3\pm0\fdg5$, $\vartheta = 327\fdg7\pm0\fdg7$, and $v_{\rm sys} = 5100\pm10$ m s$^{-1}$ (assuming $d = 162$ pc; \citealt{gaia_dr3}).  The {\tt Layer\,\,2} extractions and inferences were performed as described in Section \ref{sec:layer2} for $^{12}$CO and $^{13}$CO separately (using the same {\tt Layer\,\,1} draws).  The results are shown together in Figure \ref{fig:MWC480_layers123}: the inferred emission surfaces are comparable to those found by \citet{law21}.  

The brightness temperature profiles extracted along those surfaces in {\tt Layer\,\,3} are also shown in Figure \ref{fig:MWC480_layers123} (as was also seen in Figure \ref{fig:temp_53}, the $T_b$ dispersions are small).  We were unable to find a two-dimensional temperature prescription that satisfactorily accounts for these profiles.  For reference, the \citet{law21} best-fit model profiles are overlaid as well, and it is apparent that they have substantial deviations (particularly at larger $r$) indicative of an imperfect model prescription.  Until some insights on a more flexible $T(r, z)$ parameterization are available (a weakness we noted in Section \ref{sec:caveats}), we opted for a modified approach.  For each \{$\theta_{\tt 1}$, $\theta_{\tt 2}$\} draw, we fit exponentially tapered power-law temperature profiles to the extracted $T_b(r)$ along the surface for each line and adopted the midplane temperature profile derived by \citet{zhang21} (as approximated with uncertainties by \citealt{sierra21}): the medians for each of those three profiles are marked in the middle panel of Figure \ref{fig:MWC480_layers123}.  The $T(r, z)$ distribution was then approximated using bilinear interpolation between those three layers, assuming the vertical temperature profile saturates (becomes constant) above the $^{12}$CO surface.  The median $T(r, z)$ approximated in this way in {\tt Layer\,\,3} is shown in the bottom panel of Figure \ref{fig:MWC480_layers123}.  

\begin{figure*}[ht!]
    \centering
    \includegraphics[width=\linewidth]{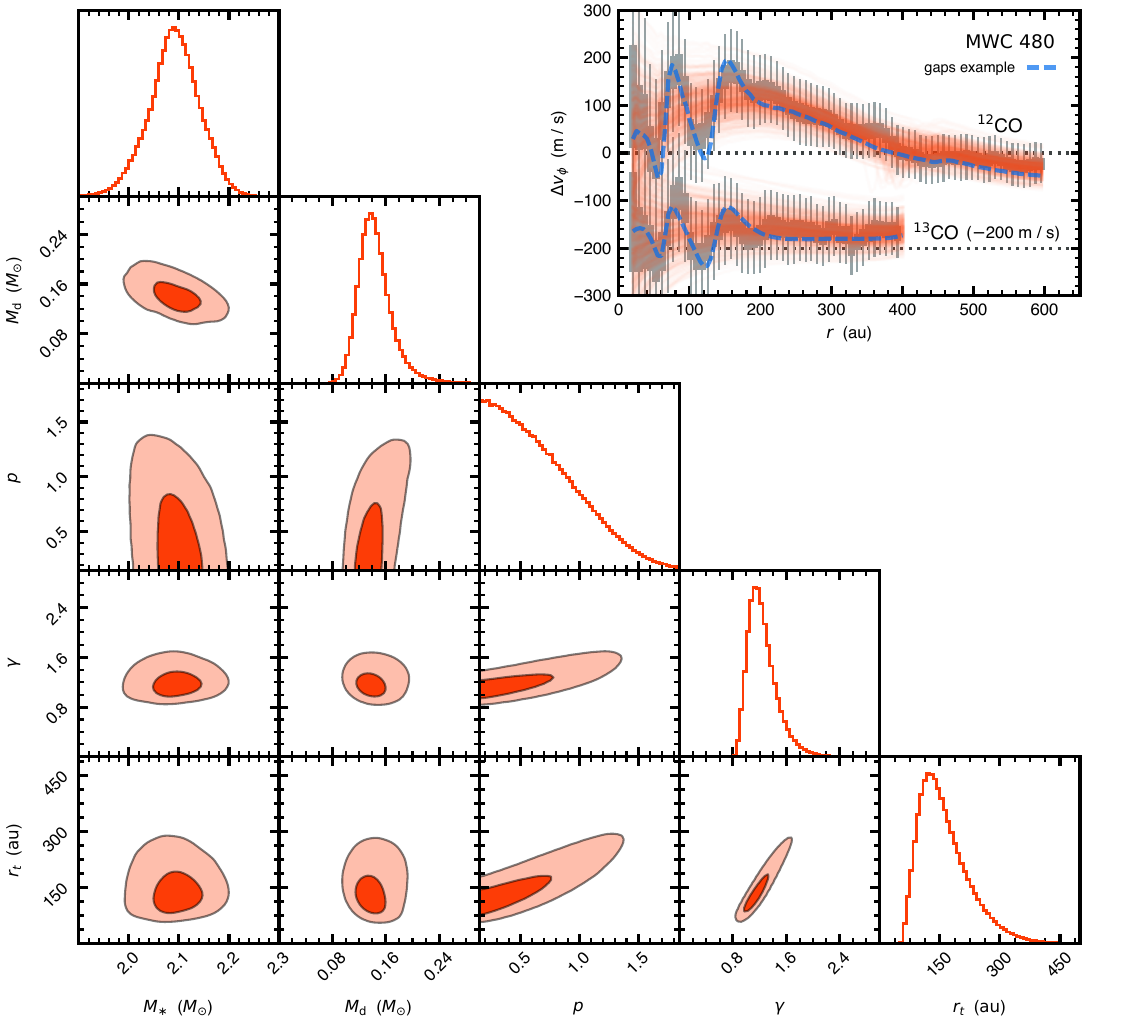}
    \caption{The marginalized $\theta_{\tt 4}$ posteriors for the joint inference of the $^{12}$CO and $^{13}$CO kinematics in the MWC 480 disk: contours mark the 1 and 2$\sigma$ confidence intervals (see Table \ref{table:mwc480}).  The panel in the upper right shows binned summaries of the extracted (residual) velocity profile data, compared with the reconstructured $\Delta v_{\sf m}$ from 1,000 draws of the $\theta_{\tt 4}$ posteriors.  Here, the residual profiles are defined with reference to the best-fit $v_\ast$ for each line.  Note that the $^{13}$CO extractions and models have been offset lower by 200 m s$^{-1}$ for clarity (as marked by the second horizontal dotted line).  The dashed blue curves show an example variant of the best-fit model that imposes two gaps on the baseline surface density profile, to demonstrate that such substructures can account for the kinematic oscillations found within $r \approx 150$ au without introducing any significant bias. \\ }
    \label{fig:layer4_MWC480}
\end{figure*}

The velocity profiles for each of 1,000 \{$\theta_{\tt 1}$, $\theta_{\tt 2}$\} draws were extracted for both the $^{12}$CO and $^{13}$CO lines as described in Section \ref{sec:layer4}.  \edits{These profiles were then corrected at smaller $r$ following the procedure detailed in Section \ref{sec:res_bias}, again using mock cubes generated with {\tt csalt}: the correction factors $\mathcal{C}_v(r)$ are similar to those shown in Figure \ref{fig:vphi_corr}}.  We set the annular spacing of these extractions to 0\farcs05, and limited consideration to the ranges $r \le 600$ and 400 au for $^{12}$CO and $^{13}$CO, respectively, based on the behaviors of their surfaces and $T_b(r)$ (Figure \ref{fig:MWC480_layers123}).  Using draws from the $T(r, z)$ prescription described above, we then performed the {\tt Layer\,\,4} inference for {\it both lines} jointly (using the sum of log-likelihoods for $^{12}$CO and $^{13}$CO lines as a composite log-likelihood).  The only other modification from the procedure outlined in Section \ref{sec:final_inference} was to adjust the soft upper boundary of the $r_t$ prior (Eq.~\ref{eq:prior4}) from 300 to 600 au, to accommodate the larger extent of the MWC 480 disk.  

The inferred $\theta_{\tt 4}$ covariances and marginalized posterior distributions are shown in Figure \ref{fig:layer4_MWC480} (autocorrelation times were $\sim$150 steps).  As before, we also show comparisons between the extracted $\Delta v_\phi(r)$ distributions and models reconstructed from posterior draws.  In this case (since we do not know $v_\ast^\ast$), we defined the residual velocity profiles with respect to the $v_\ast$ evaluated at the best-fit (peak of the marginal posteriors) for $M_\ast$ and the posterior medians of $\theta_{\tt 2}$ for each line: note that the $^{13}$CO data and models have been offset (lower by 200 m s$^{-1}$) to minimize overlap for the sake of clarity.  Summaries of these inferences are provided in Table \ref{table:mwc480}.

\begin{deluxetable}{c c c c c}[!t]
\tablecaption{Inferred MWC 480 Parameters \label{table:mwc480}}
\tablehead{
\colhead{$M_\ast$ ($M_\odot$)} & \colhead{$M_d$ ($M_\odot$)} & \colhead{$p$} & \colhead{$\gamma$} & \colhead{$r_t$ (au)}
}
\startdata
2.09 $^{+0.06}_{-0.04}$ & 0.13 $^{+0.04}_{-0.01}$ & 0.5 $^{+0.5}_{-0.4}$ & 1.1 $^{+0.3}_{-0.1}$ & 124 $^{+99}_{-21}$ \\
\enddata
\tablecomments{Uncertainties represent the 68.3\%\ (1$\sigma$) confidence intervals.  Note that the $p$ inference is formally consistent with the assumed prior (Eq.~\ref{eq:prior4}).}
\end{deluxetable}

Overall, the quality and general behavior of these posterior inferences are quite similar to those determined in the controlled injection/recovery experiments described in Section \ref{sec:results} for similarly massive disks, particularly when considering that the MWC 480 disk is larger (more radial dynamic range is available) and we have combined two spectral line tracers.  For reference, we inferred a mass ratio $M_{\rm d} / M_\ast = 7\pm1$\%\ in this case, consistent with expectations based on a comparison with Figure \ref{fig:mratio}.  We should also note that inferences using only the $^{12}$CO or $^{13}$CO data alone are consistent with the joint inference, albeit with poorer precision and relatively small biases (the $^{13}$CO data alone are essentially insensitive to $r_t$, and thereby $p$ and $\gamma$, in part due to the limited spatial dynamic range and $v_\phi$ precision).  

But despite the general quality of the inferences in this case, the simple prescriptions for the physical parameters that we have adopted are unable to explain a variety of details in the observed quantities.  Specifically, there are clear, smaller-scale ``oscillations" in the line emission distributions throughout the MWC 480 disk, identifiable in the extracted profiles for the emission surfaces, brightness temperatures, and velocities (\citealt{law21,maps3,teague21}; Figures \ref{fig:MWC480_layers123} and \ref{fig:layer4_MWC480}).  While it is not clear what physical mechanisms produce these features, they are presumably indications of localized substructures in the physical conditions and/or dynamics.  The focus here is not in accounting for these features by significantly expanding the model prescription(s), but it is worth considering how these substructures might bias the simplistic $\Sigma(r)$ inferences described above. 

\begin{figure}[ht!]
    \centering
    \vspace{0.2cm}
    \includegraphics[width=\linewidth]{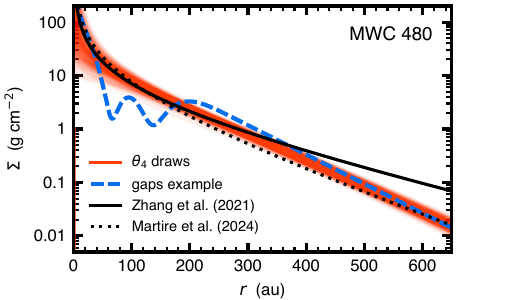}
    \caption{The surface density profiles reconstructed from random draws of the $\theta_{\tt 4}$ posteriors for the joint fit of the $^{12}$CO and $^{13}$CO kinematics in the MWC 480 disk, along with the example model that was manually tuned with Gaussian gaps to reproduce the kinematic substructures.  These profiles are compared with the $\Sigma(r)$ derived by \edits{\citet{martire24} for a similar kinematic analysis (but with fixed $p$ and $\gamma$), and by} \citet{zhang21} from independent, complementary thermo-chemical modeling of continuum and spectral line (CO isotopologue) intensities.}
    \label{fig:sigma_MWC480}
\end{figure}

An inspection of the residual velocity profiles in Figure \ref{fig:layer4_MWC480} suggest that the most significant concerns along these lines are the strong ($\pm$100 m s$^{-1}$) kinematic oscillations present for both spectral lines inside $r \approx 150$ au.  If those oscillations are intrinsically asymmetric enough (in amplitude) that the smooth model prescriptions we adopted do not recover well the mean behavior of the underlying profiles, we could expect these substructures to induce a bias in the $M_\ast$--$M_{\rm d}$ degeneracy that has characteristics similar to the resolution  problem described in Section \ref{sec:res_bias}.  As a test to guide expectations, we considered a simplistic modification of the smooth $\Sigma(r)$ in Eq.~(\ref{eq:Sigma}) that imparts these kinds of kinematic oscillations through (primarily) the associated pressure modulations (i.e., manifested through $\varepsilon_p$).  This involved introducing two gaps in the baseline $\Sigma(r)$ structure (Table \ref{table:mwc480}), centered at 65 and 135 au with Gaussian widths (standard deviations) and (fractional) depletions of 30 and 40 au and 94 and 88\%, respectively.  Those simple, manually tuned features generate kinematic oscillations that reproduce quite well the extracted $\Delta v_\phi$ profiles, as shown by the dashed blue curves in Figure \ref{fig:layer4_MWC480}.  This match suggests that these substructures are not a source of significant bias here: the constraints on $\Sigma(r)$ are dominated by the self-gravity contributions at larger $r$.  

Figure \ref{fig:sigma_MWC480} shows the $\Sigma(r)$ profiles reconstructed from random draws of the $\theta_{\tt 4}$ posteriors (as shown in Figure \ref{fig:layer4_MWC480}) and this example model with localized density gaps.  Overlaid are the $\Sigma(r)$ estimates from \edits{\citet{martire24}, based on a similar kinematic analysis (but with fixed density gradients $p = \gamma = 1$), and} \citet{zhang21}, derived from detailed thermo-chemical modeling based on the emission {\it intensities} from the dust continuum and CO isotopologue spectral lines.  The general agreement between these independent techniques is striking, particularly when considering that \citet{zhang21} relied on extrapolation beyond $\sim$400 au where the CO isotopologue (and dust continuum) emission is unavailable.  This consistency again points to opportunity for future work, where leveraging both approaches together could offer an even more robust approach for constraining the physical conditions in these gas disks.

Even though the example model $\Sigma(r)$ with gaps is not an optimized or unique solution, it is worth noting that the behavior it predicts is also generally consistent with other available diagnostics.  For example, the $\sim$65 au (0\farcs4) gap and corresponding 100 au (0\farcs6) peak are also seen as analogous substructures in the millimeter continuum \citep{long18,yliu19,sierra21} and various other emission lines (\citealt{maps3}; e.g., the $^{13}$CO and C$^{18}$O profiles shown in Figure \ref{fig:MWC480_layers123}).   This suggests there are exciting opportunities to better leverage these kinematic deviations with various intensity tracers to quantify the physical conditions associated with substructures, and particularly for characterizing localized pressure maxima and particle traps \citep[e.g.,][]{rosotti20,yen20}.

\section{Summary} \label{sec:summary}

We conducted a methodological study for a technique that relies on kinematic information from resolved spectral line observations to measure the mass and density distribution of a protoplanetary gas disk.  We found:
\begin{itemize}
    \item A layered analysis framework that sequentially extracts information from the observations and estimates relevant parameters can be used to effectively propagate uncertainties in the disk geometry, emission surface location, and physical conditions into inferences of the density distribution;
    \item The quality of constraints on \mdisk\ are most significantly affected by the ambiguities (degeneracies) in the shape of the surface density profile -- especially the taper at large $r$ -- and the quality of estimates for extrinsic (geometric) parameters; 
    \item For sensitive data with sufficiently high resolution, and adopting reasonable priors, we can recover $\Sigma(r)$ with remarkable fidelity for massive disks ($M_{\rm d} / M_\ast \gtrsim 5$\%) using only the kinematic information from a single spectral line.  The \mdisk\ precision could potentially be improved with some tool refinements and by combining tracers.
    \item We illustrated this approach using archival ALMA observations of the MWC 480 disk, inferring a gas mass ($M_{\rm d} = 0.13\,^{+0.04}_{-0.01}$ $M_\odot$, or $M_{\rm d} / M_\ast = 7\pm1$\%) and surface density profile in good agreement with a thermo-chemical modeling analysis of the spectral line and continuum intensities by \citet{zhang21}.  We also demonstrated that conspicuous kinematics features could be explained with localized density depletions at 65 and 135 au (and a corresponding pressure maximum around 100 au), consistent with the substructures identified in other (intensity-based) tracers.  
\end{itemize}
While the effects are subtle and observations are challenging, this approach for dynamical estimates of protoplanetary disk masses -- particularly at the higher end of the mass distribution, where intensity-based approaches will struggle with optical depth effects -- should have an especially promising future role in expanding our quantitative understanding of the planet formation process.

\acknowledgments We are grateful to Cristiano Longarini, Kees Dullemond, David Wilner, and Christophe Pinte for their thoughtful suggestions, to Sylvain Korzennik for his patient advice on computing, and to Feng Long and Anibal Sierra for sharing results.  \edits{We acknowledge an exceptionally helpful report from an anonymous reviewer.}  Some of the computations used in this research were conducted on the Smithsonian High Performance Cluster (SI/HPC), Smithsonian Institution (\url{https://doi.org/10.25572/SIHPC}).  S.A.~acknowledges support from the National Aeronautics and Space Administration under grant {\tt 17-XRP17\char`_2-0012} issued through the Exoplanets Research Program.  

This paper makes use of the following ALMA data: ADS/JAO.ALMA\#2018.1.01055.L. ALMA is a partnership of ESO (representing its member states), NSF (USA) and NINS (Japan), together with NRC (Canada), MOST and ASIAA (Taiwan), and KASI (Republic of Korea), in cooperation with the Republic of Chile. The Joint ALMA Observatory is operated by ESO, AUI/NRAO and NAOJ.  The National Radio Astronomy Observatory is a facility of the National Science Foundation operated under cooperative agreement by Associated Universities, Inc.

\software{
\\
{\tt analysisUtils} \citep{analysisUtils},
{\tt astropy} \citep{astropy, astropy_old, astropy_oldest},
{\tt bettermoments} \citep{bettermoments1,bettermoments2},
{\tt CASA} \citep{CASA22}, 
{\tt cmasher} \citep{cmasher},
{\tt corner} \citep{corner},
{\tt csalt} (Andrews et al.,~{\it in preparation})\footnote{\url{https://github.com/seanandrews/csalt}},
{\tt disksurf} \citep{disksurf},
{\tt eddy} \citep{eddy},
{\tt emcee} \citep{foreman-mackey13},
{\tt frank} \citep{jennings20},
{\tt gofish} \citep{gofish},
{\tt numpy} \citep{numpy}, 
{\tt matplotlib} \citep{matplotlib}, 
{\tt RADMC-3D} \citep{dullemond12},
{\tt vis\char`_sample} \citep{loomis18}\footnote{\url{https://github.com/AstroChem/vis_sample}}.
}

\clearpage

\appendix

\begin{figure}[h!]
    \centering
    \includegraphics[width=\linewidth]{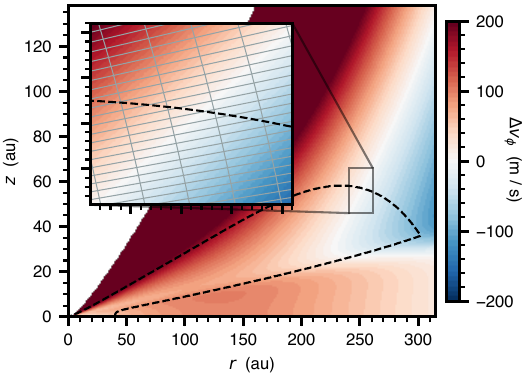}
    \caption{The main panel shows the same velocity residuals map for {\sf model B} as in Figure \ref{fig:dv} (bottom row, middle column), although over a slightly extended vertical domain.  The inset panel shows a small subset of that distribution with the locations of the {\tt RADMC-3D} grid cells (spherical coordinates) overlaid.  The inset is shown with equal aspect ratio in cylindrical coordinates, with a $\Delta v_\phi$ colorscale condensed by a factor of 5 compared to the main panel (i.e., the full colorscale represents $\pm$40 m s$^{-1}$) for clarity.  This example illustrates that the radiative transfer grid samples well the physical properties of the model near the top of the CO-rich layer, where most of the line emission is generated.}  
    \label{fig:gridding}
\end{figure}

\section{Notes on the Model Gridding} \label{app:grid}

In generating synthetic datasets using {\tt RADMC-3D} as described in Section \ref{sec:RT}, the physical conditions, kinematics, and molecular abundances for a given model were specified on a fixed grid in spherical coordinates ($R$, $\Theta$).  Here $R = \sqrt{r^2 + z^2}$ is the distance from the origin (stellar position), and $\Theta = \arctan{(r/z)}$ is the polar angle (such that $\Theta = \pi / 2$ is the midplane).  This coordinate frame has distinct numerical advantages for flared disk geometries, since it effectively has a built-in refinement to cover an appropriate dynamic range in height at all radii.  That said, it can be somewhat clumsy for defining physical structures (see below).  We defined the grid to have 360 cells in the $R$ dimension, distributed logarithmically from 1 to 500 au.  In the polar angle dimension, we distributed 240 cells logarithmically from the midplane (denser) to the pole (sparser).  Figure \ref{fig:gridding} shows an example of the grid sampling for the {\sf model B} velocity residuals, demonstrating that the physical properties are well-sampled in the relevant model regions.  

Some attributes of a model can be directly calculated on this grid, simply using the ($r$, $z$) $\rightarrow$ ($R$, $\Theta$) mapping (e.g., $T$ or $v_\ast$).  However, for other quantities that rely on integrals in cylindrical coordinates (e.g., $\rho$ or $\varepsilon_g$), we computed a refined (one-dimensional) subgrid in cylindrical coordinates around each cell in the {\tt RADMC-3D} grid.  This is notably inefficient, but it ensured that numerical artifacts from interpolation methods did not introduce any undesired bias into the synthetic data products (and we only needed to compute a small number of datasets for the purposes of this article).  Some testing verified that the pressure gradient in the $\varepsilon_p$ kinematic deviation (Eq.~\ref{eq:bulk_motion}) could be calculated in a straightforward way (once $\rho$ was determined), using the chain rule,
\begin{equation}
    \frac{\partial P}{\partial r} = \frac{\partial P}{\partial R} \sin{\Theta} + \frac{1}{R}\frac{\partial P}{\partial \Theta} \cos{\Theta},
\end{equation}
without resorting to another subgrid calculation.

\section{MWC 480 Geometric Parameters} \label{app:mwc480_layer1}

\begin{figure}[b!]
    \centering
    \includegraphics[width=\linewidth]{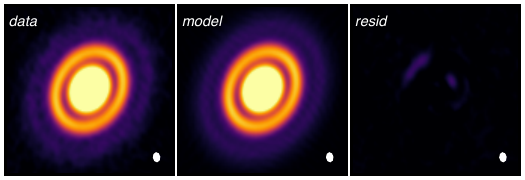}
    \caption{The images synthesized from the observed 225 GHz continuum visibilities for the MWC 480 disk ({\it left}), the best-fit {\tt frank} model ({\it middle}), and the associated residuals ({\it right}) on the same spatial (425 au on a side) and intensity (an {\tt asinh} brightness temperature stretch from 0 to 5 K) scales.  The synthesized beam dimensions ($0\farcs16 \times 0\farcs11$) are shown in the lower right corners of each panel.}  
    \label{fig:mwc480_contfit}
\end{figure}

For the analysis described in Section \ref{sec:mwc480}, we derived the geometric parameters in {\tt Layer\,\,1} from the visibility measurements of the $\sim$225 GHz continuum associated with the same ($^{12}$CO and $^{13}$CO) spectral line observations \citep{maps1}.  Because those data share the same calibration procedure, their relative offsets from the phase center ($\Delta x$ and $\Delta y$) must be the same.  Presuming there is no warp or large-scale axisymmetry, the projection angles ($i$ and $\vartheta$) should also be the same.  We followed the procedure outlined by \citet{andrews21}.  To briefly summarize, for a given choice of \{$\Delta x$, $\Delta y$, $i$, $\vartheta$\} we deprojected the continuum visibilities and modeled them assuming axisymmetry with the {\tt frank} package \citep{jennings20}.  That process was iterated for different geometries until we reached a satisfactory map that minimized the imaged residual visibilities.  Figure \ref{fig:mwc480_contfit} shows the images generated from the data, an optimized model, and the corresponding residual visibilities on the same spatial and intensity scales.

\begin{figure}[h!]
    \centering
    \includegraphics[width=\linewidth]{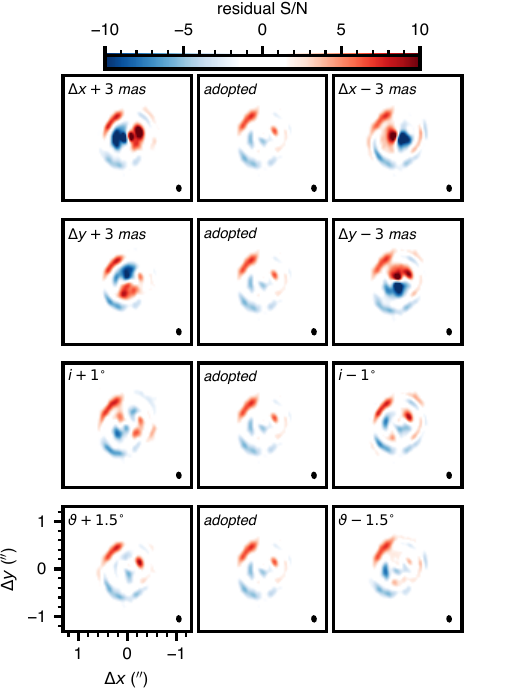}
    \caption{The images synthesized from the residual 225 GHz continuum visibilities for the MWC 480 disk for the adopted geometry -- with $\Delta x = 5$ mas, $\Delta y = 1$ mas, $i = -36\fdg3$, $\vartheta = 327\fdg7$ -- are shown in the middle column (same for all rows).  Each row compares a variation in a single parameter, as annotated.  The offsets introduce strong residuals (here showing only half the standard deviation adopted in the inferences), while the projection angles have more subtle effects (the $i$ and $\vartheta$ ranges are $\pm$2$\sigma$ for the adopted ranges).  All plots are shown on the same spatial scale (as in Figure \ref{fig:mwc480_contfit}) and using a linear stretch of the residual S/N (normalized by the RMS = 31 $\mu$Jy beam$^{-1}$).  The beam dimensions are marked in the lower right of each panel.}  
    \label{fig:mwc480_contalt}
\end{figure}

The {\tt Layer\,\,1} extractions assumed uncertainties based on a qualitative assessment of the outcomes of this procedure over a modest-sized grid of geometric parameters.  Figure \ref{fig:mwc480_contalt} illustrates the estimated uncertainties in terms of the imaged residuals (here in S/N units, with respect to the mean RMS noise value of 31 $\mu$Jy beam$^{-1}$) for positive and negative variations of each parameter around the adopted values, with all other geometric parameters fixed.  The bright, centrally peaked emission distribution means that the formal precision of the offset parameters, $\Delta x$ and $\Delta y$, is especially high: the residuals are notably stronger for deviations of only $\sim$1 mas in either dimension.  However, we conservatively inflated the adopted uncertainties (to $\pm$7 mas) because of the lingering asymmetric residuals we find even in the optimized case.  The inclination and position angle variations are more subtle, but still exhibit the distinctive patterns in the residual images noted by \citet{andrews21}.  Even for our best estimates, there are substantial residuals.  The strongest of these are primarily associated with the bright emission ring along the minor axis, consistent with expectations for an elevated (and presumably optically thick) emission surface.  

The adopted geometry is similar to those estimated by \citet{teague21} and \citet{izquierdo23} from the same data.  We found $i$ closer to the latter estimates based on the rare isotopologues, and consistent $\vartheta$.  But quantitatively, the offsets are demonstrably different compared to both studies (and as noted above, are constrained well): even though the shifts are still well within the size of the synthesized beam, these differences can have non-trivial effects on the velocity extractions.  As noted above, this can be one of the important systematic limitation for kinematic studies.

\clearpage
 
\bibliography{references}

\clearpage

\end{document}